\documentclass[
11pt,
superscriptaddress,
showkeys,
showpacs,
nofootinbib
]{revtex4-1}

\usepackage[utf8]{inputenc}
\usepackage[english]{babel}
\usepackage{amsmath}
\usepackage{amsfonts}
\usepackage{amssymb}
\usepackage{graphicx}

\usepackage{color}

\usepackage{hyperref}   % use for hypertext links, including those to external documents and URLs
\newcommand{\melec}{m_{\mathrm{e}}}
\newcommand{\ZZ}{\mathbb{Z}}
\newcommand{\URL}[1]{url: \url{#1}} %\nolinkurl{}
\newcommand{\DOI}[1]{doi: \href{http://doi.org/#1}{#1}}

\begin{document}

\title{The Traveling-Wave Tube in the History of Telecommunication}
% Force line breaks with \\

\author{Damien~F.~G.~Minenna}%
 \email[Electronic address: ]{damien.minenna@univ-amu.fr}
\affiliation{%
Centre National d'\'Etudes Spatiales, 31401 Toulouse cedex 9, France
}%
\affiliation{%
Aix-Marseille University, UMR 7345 CNRS PIIM, \'equipe turbulence plasma, case 322 campus Saint J\'er\^ome,  av.\ esc.\ Normandie-Niemen, 13397 Marseille cedex 20, France
}%
\affiliation{%
Thales Electron Devices, rue Lat\'eco\`ere, 2, 78140 V\'elizy, France
}%
\author{Fr{\'e}d{\'e}ric~Andr{\'e}}%
 %\email[Electronic address: ]{frederic.andre@thalesgroup.com}
\affiliation{%
Thales Electron Devices, rue Lat\'eco\`ere, 2, 78140 V\'elizy, France
}%
\author{Yves~Elskens}%
 %\email[Electronic address: ]{yves.elskens@univ-amu.fr}
\affiliation{%
Aix-Marseille University, UMR 7345 CNRS PIIM, \'equipe turbulence plasma, case 322 campus Saint J\'er\^ome,  av.\ esc.\ Normandie-Niemen, 13397 Marseille cedex 20, France
}%
\author{Jean-Fran{\c{c}}ois~Auboin}%
 %\email[Electronic address: ]{frederic.andre@thalesgroup.com}
\affiliation{%
Thales Electron Devices, rue Lat\'eco\`ere, 2, 78140 V\'elizy, France
}%
\author{Fabrice~Doveil}%
 %\email[Electronic address: ]{yves.elskens@univ-amu.fr}
\affiliation{%
Aix-Marseille University, UMR 7345 CNRS PIIM, \'equipe turbulence plasma, case 322 campus Saint J\'er\^ome,  av.\ esc.\ Normandie-Niemen, 13397 Marseille cedex 20, France
}%
\author{J{\'e}r{\^o}me~Puech}%
 %\email[Electronic address: ]{damien.minenna@univ-amu.fr}
\affiliation{%
Centre National d'\'Etudes Spatiales, 31401 Toulouse cedex 9, France
}%
\author{{\'E}lise~Duverdier}%
\noaffiliation

\date{March 27, 2018. Submitted. \copyright The authors}% It is always \today, today,

\begin{abstract}
The traveling-wave tube is a critical subsystem for satellite data transmission. Its role in the history of wireless communications and in the space conquest is significant, but largely ignored, even though the device remains widely used nowadays. This paper present, albeit non-exhaustively, circumstances and contexts that led to its invention, and its part in the worldwide (in particular in Europe) expansion of TV broadcasting via microwave radio-relays and satellites. We also discuss its actual contribution to space applications and its conception. The originality of this paper comes from the wide period covered (from first slow-wave structures in 1889 to present space projects) and from connection points made between this device and commercial exploitations. The appendix deals with an intuitive pedagogical description of the wave-particle interaction.
\end{abstract}

%Use showkeys class option if keyword
    
\maketitle

\section*{Introduction}

On November~12, 2014, the space probe \textit{Rosetta}, built by the European Space Agency (ESA), detached its lander module \textit{Philae} which performed the first successful landing on a comet, more than 475~million kilometres away from the Earth. This historic achievement was met thanks to years developing critical systems of the probe, like solar cells, trajectory computer, or propulsion parts. Yet the success of a mission depends crucially on the spacecraft capacity to transmit data across the vacuum of space.
Messages must contain enough information, and must be sent with enough power to be captured on the ground, but avoid spending too much electric power, which is scarce in space.
Such communication systems from \textit{Rosetta} enabled us to receive scientific data to understand the history of the solar system as well as stunning images of the comet.

\bigskip

Telecommunication subsystems are mainly composed by an antenna, a receiver and a transmitter. Inside the transmitter, we need a device that amplifies radio-waves enough for us to communicate with spacecrafts, even at more than billions of kilometres ---like with the probes \textit{Voyager~1} or \textit{New Horizons}, discussed below. This wave amplifier meets a large number of criteria to be operational in space. For data emission, it needs to reach high signal amplitude with very low noise, and also needs a large bandwidth at high frequency corresponding to the amount of transmitted information. Lastly, this device needs to resist the shocks of its rocket launch and to operate for years in the dangers of space, in particular aggressive radiations.

All those conditions are fulfilled by the traveling-wave tube (TWT). Its apparition was followed by the expansion of long-range communications with the worldwide development of TV broadcasting since the 1950s, and multiplication of telecom satellites since the 1960s. Today, this device remains at the cutting edge, and still contributes to transmissions for major satellites and space probes.

\begin{figure}[!t]
\centering
\includegraphics[width=\textwidth]{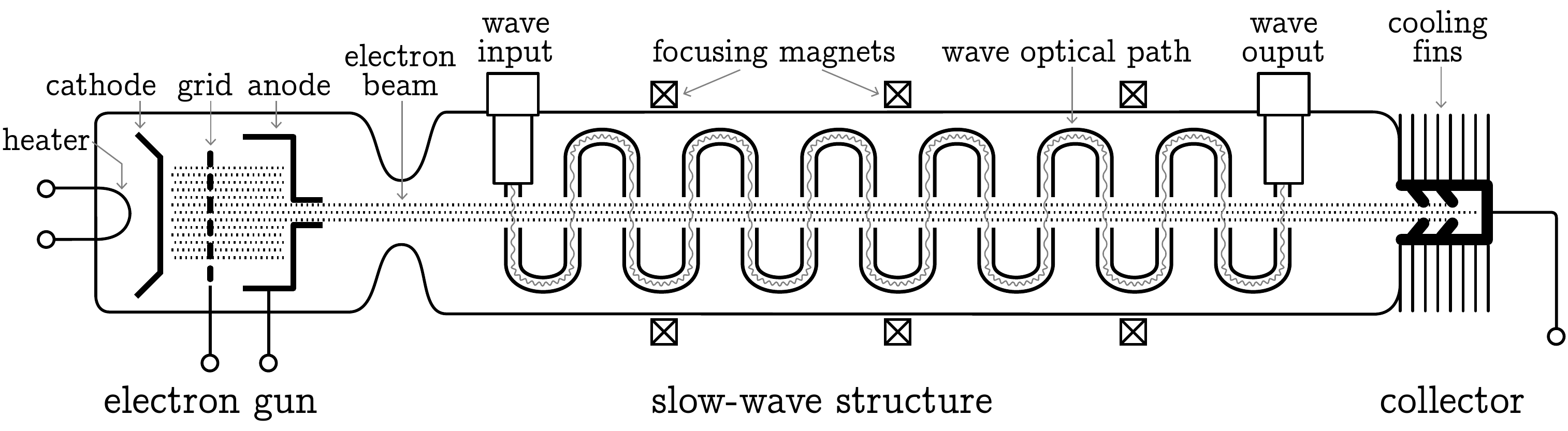}
\caption{\label{f:twt} The traveling-wave tube is a simple electrical equipment comprising three parts: An electron gun that emits the beam; a delay line ---also called a slow-wave structure--- where the signal travels and where the wave-particle interaction is performed; a collector to take back the energy of electrons and limit losses. All those parts are under an extremely high vacuum. Beam focusing magnets ---or periodic permanent magnet (PPM)--- keep electrons moving on the longitudinal axis. 
In space traveling-wave tubes, the optical path is a metallic helix, as shown in figure~\ref{f:helix}; 
the schematic here displays a folded wave\-guide. 
Space tubes are 10-30~centimetres long. For higher frequency regimes, one uses shorter tubes, but they are harder to assemble.}
\end{figure}

\bigskip

Traveling-wave tubes\footnote{In the past, they were sometimes called forward-wave tubes, or traveling-wave valves. Also, some authors write ``travelling'' in British English with a double l.} (see figure \ref{f:twt}) are electronic tubes (a.k.a.\ vacuum electron devices or thermionic valves) \cite{fai08,gil11} used to amplify radio-waves, and have a lot of applications, like radar, electronic warfare, television or radio broadcasting, Internet, sending data from space probes, planes and ships transponders, GPS, and imaging devices, or even studying the scientific characterisation of plasmas. 
Like all other vacuum electron devices, it is based on the momentum transfer ---similar to kinetic energy transfer--- from accelerated electrons to a radio-frequency wave via an interaction in a vacuum environment (see appendix~A).
When combined with a power supply, TWTs are preferably called traveling-wave tube amplifiers (TWTAs). 
They are not the most powerful of vacuum electron devices, but their main features are their large bandwidth and their excellent power efficiency: ideal for long-range communications. Combined with their robustness, their high power and their long lifespan, traveling-wave tubes became quickly indispensable for space programs.

\section*{Wireless signals before 1945}

The history of wireless communications\footnote{In fact, long-range communications began a long time ago with smoke signals, drums, whistled languages, beacons, semaphore, etc.} started long before the invention of the traveling-wave tube.  It all begun in 1888, when the German physicist Heinrich R.~Hertz observed experimentally the propagation of electromagnetic waves in the air for the first time \cite{her88}, twenty years after they were theorized by the Scottish mathematician and physicist James Clerk Maxwell \cite{max65}. 
%Legend has it that Hertz would have said to his students, after they questioned him about the interest of those Hertzian-waves, that this was just a scientific curiosity without any application possible. 
Six years later, the Italian Guglielmo Marconi built one of the first radio transmitters with a range of a hundred meters. With his Wireless Telegraph \& Signal Company, Marconi realized a cross-Channel radio-telegraphic transmission in 1899, then a transatlantic emission in 1901 \cite{coe61}. At this time, all intercontinental telecommunications were performed thanks to submarine telegraph cables \cite{sch08}. Marconi's ambition was to build a global wireless network to rival with cables. He started to open communication stations, but he immediately faced competition when electrical equipment manufacturers and some countries started their own developments. 
Indeed, to help the German navy, Siemens and the Allgemeine Elektricit{\"a}ts-Gesellschaft (AEG) were encouraged to create the radio company Telefunken in 1903.
That same year, Gustave A.~Ferri{\'e} performed long-distance radio experiments\footnote{Unintentionally, Ferri{\'e} experiments granted an important usefulness to the Iron Lady and protected it from its scheduled demolition. The Eiffel Tower became and still is an important radio station. Before him, Eug{\`e}ne Adrien Ducretet performed, in 1898, a sound emission by wireless waves between the Eiffel tower and the Panth{\'e}on, 4~km away \cite{eil00}.} at the top of the Eiffel tower \cite{fer11}.
Thereafter, in 1915, the first transatlantic telephony call is achieved by AT{\&}T\footnote{One of the pioneers of telephony, the Scot Alexander Graham Bell, founded the Bell Telephone Company in 1877. The company became the American Telephone \& Telegraph Company (AT{\&}T) in 1885, and was at times the world's largest telephone company.}, between Arlington, Virginia and the Eiffel Tower, followed by a call between Arlington and Honolulu.

%CSF do not to be confuse with the Belgian Marconi subsidiary; the Compagnie de T{\'e}l{\'e}graphie sans Fils (CTSF), founded in 1901 to create a wireless line between Belgium and Congo; nor than be confuse with the French Marconi subsidiary; the Compagnie Maritime et Coloniale de T{\'e}l{\'e}graphie sans Fils, founded in 1903.

Marconi's first successful transatlantic message, sent out more than 3500~km away, had raised some interrogations regarding the curvature of the Earth's surface. Indeed, it would require two towers 550~km tall (or one 900~km tall) to have a direct visual connection between both stations. Actually, the Earth's ionosphere plays an important role because it reflects waves with a frequency lower than 300~MHz, wiping out limitations due to our planet curvature, thus allowing long-range emissions. Reflections occurred in one of the layers of the atmosphere composed of ions called the Kennelly-Heaviside layer (100~km above the ground) and theorized separately by Arthur E.~Kennelly \cite{ken02}, and Oliver Heaviside \cite{hea02}.
It was thanks to this property that the British Broadcasting Corporation (BBC) was able to broadcast radio from London over Europe during World War II, without any facilities on the Old Continent. But this application is extremely sensitive to weather conditions.

\bigskip

On the other hand, the history of vacuum electron devices started in the late 19\textsuperscript{th} century with the discovery of thermionic emission ---the electron flux emission coming from heated metal filaments--- utilized for incandescent light bulbs\footnote{The same kind of light bulbs used for domestic consumption during the 19\textsuperscript{th} and 20\textsuperscript{th} centuries.}. In 1904, the English physicist Sir John A.~Fleming, while working for the Marconi company, used this effect to detect radio waves and built the first vacuum tube: the Fleming valve (a.k.a.\ the vacuum diode) \cite{Fle05}.
But the first practical tube is credited to the American Lee de Forest when he built, in 1906, a triode (named ``audion''), which was able to better receive, radiate and amplify electromagnetic signals \cite{def08}.
As electromagnetic waves were also reflected by metallic surfaces, as proved by Hertz, another major application of vacuum tubes was radar (RAdio Detection And Ranging) operations.  Patents about radars were filed during the 1900s, but it is only in the 1930s and the 1940s that they began to gain in importance for military applications. Long-range detection of small objects (boats or planes) requires to broadcast strong electromagnetic powers that only magnetrons ---another kind of vacuum electron device--- were able to generate at the moment. 
During the Second World War, radars made thanks to the magnetron were one major element of the Allies' victory.

\bigskip

Between the two World Wars, television broadcasting begun timidly to emerge to the general public. For instance, RCA\footnote{The Radio Corporation of America (RCA) was formerly the American Marconi, a subsidiary of the British Marconi Company before 1919.} is known to have broadcast \cite{gol46} the first TV programs for New Yorkers on April~30, 1939, from the top of the Empire State Building in New York City and from a series of relay stations spanning the length of Long Island.
But instead of telegraphy and low definition radio, applications for a large number of telephone calls simultaneously or TV emissions need larger flows of information sent, requiring wireless signals with higher frequencies.
And for those higher frequencies (above 300~MHz), the ionosphere does not reflect the waves\footnote{In fact, microwaves (from 300~MHz to 300~GHz) are slightly reflected by the troposphere (15~km).  This effect, discovered in the 1950s \cite{boo50}, was used to increase the range of radio-relays, but it needed powerful amplifiers and huge antennas.} any more, imposing the use of relays for line-of-sight propagation to compensate for the Earth curvature. Moreover, the range and the noise of amplifiers limited the expansion of telephones and TV by Hertzian-waves, until the appearance of better devices.

\section*{Inventions of the traveling-wave tube}

Fatherhood of the traveling-wave tube (TWT), as we know it today, is most often granted to an Austrian refugee, Rudolf Kompfner in 1942 ---notably after his public announcement in 1946--- when he was secretly working on microwave vacuum tubes for the British Admiralty at the University of Birmingham during World War II.
But the history of this device is more complex because the traveling-wave tube was, consecutively, discovered thrice independently.
In fact, two fundamental concepts were needed to conceive a TWT: the slowing-down of  electromagnetic waves, and the addition of an electron beam inside the tube to have the wave-particle interaction.

\bigskip

\begin{figure}[!t]
\centering
\includegraphics[width=0.4\textwidth]{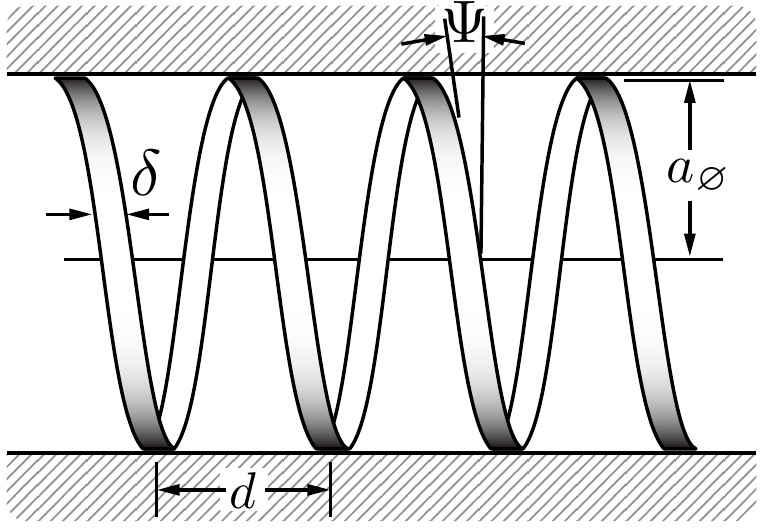}
\caption{\label{f:helix} The slow-wave structure is used to elongate the optical path of waves. For a helix, waves are slowed down by a factor $2 \pi a_{\varnothing} / d$. In space tubes, this represents generally one-tenth of the speed of light. The geometry of the slow-wave structure is chosen to match the wave velocity with the electron speed: there is resonance, and some momentum of the electrons can be transferred to the wave for amplification.}
\end{figure}

One year after his extraordinary discovery that electromagnetic waves can propagate in the air, Hertz, scrupulous to get on with his studies, did the earliest work \cite{her89} on electromagnetic wave-guides: structures that confine and guide waves.
Investigating the velocity of waves, he realized they could be steered along metallic guides.
Because of the increase of their optical path, waves will take more time to cross the guide than if they were going straight, hence Hertzian-waves are slowed down, and such frameworks are called slow-wave structures or delay-lines. 
Hertz built the first metallic helical wave-guide in 1889: ``I [Hertz] rolled a wire 40~metres long into a spiral 1~cm in diameter, and so tightly that the length of the spiral was 1.6~metre'' \cite[p.~158]{jon93}. It was a simple coil where waves were 5~times slower along the guide longitudinal axis.
But the slow-wave structure alone cannot provide any wave amplification without an electron beam.
Also, Hertz' pioneer achievement on wave-guides was paltry compared to the theoretical works performed by three English physicists: Sir Joseph J.~Thomson \cite{tho93}, Lord Rayleigh \cite{ray97}, and Henry C.~Pocklington \cite{poc97}.
They have been followed by few other authors early in the 20\textsuperscript{th} century,  but only the development of traveling-wave tubes in the 1950s increased the interest for the helical structure. 
Other forms of slow-wave structures were also designed in parallel with the development of earliest radio sets, mostly in the 1930s, when klystrons and magnetrons, two kind of vacuum electron tubes, appeared. %Indeed, the simplest radio system is composed of an emitter/receiver and an antenna linked by a wave-guide. %another example is the coaxial cable between a TV and a parabolic dish. 
However, the helix guides (see figure~\ref{f:helix}) still remain the most used structure for space TWTs because they allow broader bandwidth than others, meaning higher flows of information.

To build a traveling-wave tube, we need another fundamental idea: adding an electron beam to the slow-wave structure in a vacuum tube, which enables the wave amplification.
This idea was recognized in 1933 by the Russian Andrei V.~Haeff, a young researcher of the California Institute of Technology (Caltech), inspired after watching surfers on Santa Monica Beach, California; he concluded that the surfboard speed and the wave velocity had to match so the surfer can properly use the wave energy \cite{Cop15a}.
In his first patent \cite{hae33}, Haeff described a device ``for generating, controlling and measuring extremely high frequency waves'', to which he refers sometimes by the term ``travelling waves''. 
His device used two parallel helical slow-wave structures with an electron beam flowing between them: it was the first traveling-wave tube, even though its slow-wave structure looks unusual.
It seems that Haeff used his TWT to build a portable radio transmitter and receiver operating at 750~MHz.
But at this time, electron guns were not good enough to provide an efficient focusing of the beam, so his device was not very efficient.
The next year, Haeff joined the Radio Corporation of America (RCA) at Harrison, New Jersey, and sold them his patent, but the corporation did not permit Haeff to develop his invention further, and his discovery was largely ignored.

\bigskip

Meanwhile in the Netherlands, the traveling-wave interaction principle was the first time appreciated theoretically by Klaas Posthumus at the Philips NatLab\footnote{To avoid depending on third-party patents, Gerard and Anton Philips established in 1914 their company's own ``Physics Laboratory'', which turned into a world class fundamental and applied research facility, growing to over 2000 employees in the 1970s. Breakthroughs and achievements at the Nat(uurkundig) Lab(oratorium) include nonlinear dynamics (Balthasar van der Pol), quantum physics (Hendrik Casimir), magnetic resonance imaging, audio cassettes and the compact disc \cite{phi14}.}  (Eindhoven) when investigating the newly-invented magnetron \cite{pos35}.
He found that the axial component of the rotating wave velocity was synchronised with the average speed of electrons. As a result, the electron energy was transferred to the wave and amplified it.

The variety of early microwave devices built before 1945 is astonishing, and some inventors came very close to make a TWT. 
The American researcher Frederick B.~Llewellyn, working at Bell Labs\footnote{The Bell Telephone Laboratories \cite{ger12}, now Nokia Bell Labs, were founded in 1925 by AT\&T and the Western Electric as a independent division to conduct research and development. These laboratories are famous worldwide for their numerous contributions in telephony, TV, space communications, information theory, radio astronomy, mathematics, computer science, etc., and are associated with the discovery of transistors, photovoltaic cells, CCD captors, or optical fibers. They have also employed a lot of famous award-winning researchers including eight Nobel prizes and three Turing awards.}, proposed a patent\footnote{In fact, the use of wave-electron interaction along distributed or wave circuits within one or more vacuum tubes has
been proposed in various patents \cite{per35,lle36,zwo36,pot35,rob35,stc38,stc39,cla38}.} \cite{lle40} in 1940. His amplifier was built ``to secure a useful cooperative relation between the alternating electric field within the guide and electrons traversing that field''. However, Llewellyn's invention was not really a TWT because his folded wave-guide ---a kind of delay line comparable to the one sketched by the figure \ref{f:twt}--- was closed in two spots at the middle of the tube, making his invention working more like a klystron with two (very long) cavities having some aspects of a TWT.

\bigskip

The second time the traveling-wave tube was invented was in 1940 by the Swedish research physicist Nils E.~Lindenblad when working at Rocky Point, New York. This time, the patent \cite{lin40} proposed a modified version of Haeff's tube, to put the electron beam inside through a helix slow-wave structure.
Lindenblad designed the first modern version of the helix traveling-wave tube as we know it today and introduced it as a ``device capable of efficiently amplifying a wide band of frequencies such as would be demanded by a multichannel radio relay amplifier''.
In his patent, the basics of TWTs are well understood: ``the speed of electrons [...] is made to be substantially equal to the axial velocity of propagation of the electromagnetic wave''.
Lindenblad estimated possible to amplify a frequency band from 30 to 390~MHz without any important variations over the band.
In addition, he also recognized that the pitch (the gap $d$ in figure \ref{f:helix}) of the helix can be modified to maintain synchronism with the electron beam. As electrons lose velocity inside the tube, the pitch is decreased to slow-down the electromagnetic wave, so the amplification can be performed during a longer time: a method now often used and called ``taper'' or ``pitch tapering''.

But Lindenblad's discovery is controversial because he was an antenna specialist at RCA at the same time as Haeff, and their labs were about 100~km apart, though there is no evidence that they ever met \cite{Cop15a}.
And more curious is the fact that it was the same attorney who filed both their patents for the RCA, but the one by Haeff was never cited in Lindenblad's patent of 1940. 
The lack of theoretical models about TWTs and the significantly different designs of both inventions could explain why Lindenblad did not refer to Haeff.
However, after the rise of the TWT popularity, Lindenblad filed a new patent \cite{lin47a} in June 1947, this time with a reference to Haeff's one. 
Regardless of this anecdote, early works of Haeff\footnote{Andrei Vasily Haeff is also known for his invention of another kind of vacuum tube: the inductive-output tube (IOT, a.k.a.\ the klystrode) \cite{hae39b}. These tubes built by RCA were used on 1939 to broadcast TV programs from the top of the Empire State Building in New York City \cite{gol46}.} and Lindenblad\footnote{Nils Erik Lindenblad is credited with more than 300~patents, and is mainly known for his antennas. He developed the television antenna placed on top of the Empire State Building in 1939 \cite{lin39}.} on traveling-wave tubes were never really recognized nor used, and are only mentioned in very few historical accounts \cite{wat54}, nor referred as TWTs \cite{pie47a,war56,pie62}.

\bigskip

As previously stated, the traveling-wave tube was then rediscovered a third time by Kompfner in late 1942 at Birmingham. During wartime, the limited information exchange about radio and radar technologies might explain why the patents by Lindenblad ---published in October, just a few months before--- were unknown across the Atlantic. 
When Kompfner joined\footnote{Rudolf Kompfner had a training as an architect, but he was a tinkerer with cathode-ray tubes and klystrons when he joined the Admiralty.} the Admiralty in 1941, to work on valves at the Physics Department, Birmingham University, one of his first goals was to help a team to build a klystron amplifier, which would be more sensitive ---with a stronger wave-electron interaction--- than the crystal-mixer receivers available at that time, with the intent to improve the range of radars. To do this, we can either increase the power output or boost the sensitivity of the receiver. 
After a year of working on this problem, Kompfner concluded that if the klystron was relatively inefficient and had a narrow bandwidth, it was because the device had a fundamental weakness: the coupling between the electron beam and the radio-frequency field in the resonator gaps was too weak, because electrons spent too much time in the field, losing part of the energy they had gained a little earlier, or vice-versa.
Therefore, Kompfner made the brilliant suggestion to match the wave and electrons velocities. This led him to discuss with delay-line experts and to design the TWT concept.
The next year, in November 1943, Kompfner used a helix as slow-wave structure to build his first traveling-wave tube with a modest power gain. Later he produced a TWT at a frequency of 3.3~GHz with a better sensitivity than the best klystron ever built at this time \cite{kom76}.

\bigskip

At that moment, the TWT was still a secret device developed in wartime.

\section*{Public announcement and developments}

Kompfner moved to the Clarendon Laboratory, Oxford, in 1944, where he continued his work, aiming to find a theory that would enable design optimization. 
Seeing the great opportunities given by TWTs, he was helped by more and more people, including his research assistant Joseph Hatton.
Visiting the Clarendon Laboratory at this time, John R.~Pierce, an American researcher of Bell Labs, examined the device and realized the significance of the wide bandwidth available, ideal for microwave telecommunications.
Indeed, the main purpose of Bell Labs ---as part of a telephone company--- was to develop communication systems, even though they also worked on radars during the war.
With the aim to develop a good theory about TWTs, Pierce and Kompfner organized a partnership, then Pierce brought the concept with him to the United States.

The results of British wartime investigation on the traveling-wave tubes were presented to public announcement by Hatton at the 4\textsuperscript{th} IRE\footnote{In 1963, the Institute of Radio Engineers (IRE) merged with the American Institute of Electrical Engineers to became the Institute of Electrical and Electronics Engineers (IEEE).} Electron Tube Conference at Yale University, New Haven, on June 27 and 28, 1946. Hatton described some of the British results, and later in this conference, Pierce, associated with his colleague Lester M.~Field\footnote{Lester M.~Field was employed at Bell Labs for only two years, between 1944 to 1946, after his Ph.D.\ Thereafter, he joined Stanford university, and became, in four years, its youngest professor at the age of 32. He was one of the founders of the Electronics Research Laboratory at Stanford. This lab was well known for TWT research. After a transition by Caltech, he became vice-president and associate director of the Hughes Research Laboratories (HRL) where he was involved in TWT developments and in space programs.}, revealed how far the research at Bell Labs had progressed. 
Since the IRE conferences had a good reputation among electrical engineers, the discovery became immediately famous in the sphere of vacuum electron devices, and Kompfner's name became inseparable from his invention. 
After the conference, worldwide task forces of industrial and research labs started to develop their own versions of tubes. 
For instance, in the Soviet Union, teams of young military servicemen started to better understand the electron-wave theory and to build TWTs; but due to the Cold War, their advancement stayed barely known beyond the Iron Curtain \cite{los49,cia53,pch03}. 

A few days after the announcement, \textit{The New York Times} ostentatiously reported the possibility to send ten thousands phone calls at once by ``a device that eventually may provide the means of setting up more channels for long-distance communications than they will know what to do with'' and ``it is expected to do as much for the future of very-highfrequency [sic] nation-wide communication as the deForest `audion' did for the broadcast and world-wide telephony and telegraphy pioneers'' \cite{ken46}.

In August 1946, the first general description TWTs appeared \cite{bar46}, followed the next months by articles \cite{roc46,whi46,wil46} in a few journals of electronics.
In November, Kompfner published \cite{kom46} an introduction about his invention in the journal \textit{Wireless World}\footnote{The British electronics journal \textit{Wireless World} ---very popular among amateurs and professionals of audio and electronic devices--- was originally published by the Marconi Company under the name \textit{The Marconigraph} from 1911 to 1913. Since 1984, it is renamed the \textit{Electronics World}.}, followed by Pierce the next month \cite{pie46b} in the \textit{Bell Labs Record}.
The world's first peer-reviewed publication on a TWT theory was submitted, in December 1946, by two French researchers, Andr\'e Blanc-Lapierre and Pierre Lapostolle \cite{Bla46} of CNET\footnote{The Centre National d'{\'E}tudes des T{\'e}l{\'e}communications (CNET) was an independent division of the Postes, T{\'e}l{\'e}graphes et T{\'e}l{\'e}phones (PTT) administration (nowadays Orange).}, immediately followed by Jean Bernier \cite{ber47} of CSF\footnote{The French Compagnie G{\'e}n{\'e}rale des T{\'e}l{\'e}phonies sans Fils (CSF) was founded in 1918 by Emile Girardeau ---a former member of Ferri{\'e}'s team--- to compensate the lack of French radio companies, and probably because the army bought equipments only from legal business entities. CSF became Thomson-CSF in 1968 after merging with the Compagnie Fran\c{c}aise Thomson-Houston (CFTH). CFTH was also one of the early companies to investigate TWTs \cite{rou47}.
In 2000, the group became Thales after another merger with AEG-Telefunken.
Their tube division, Thales Electron Devices (TED), is currently one of the leaders in the space TWTs market.}
in January 1947 and by Pierce and Field \cite{pie47a,pie47b} in February. 
By the end of 1950, probably a hundred papers and patents had been published about the new tube. A famous textbook \cite{pie50}, written by Pierce and still a classic, established the robust basis of the device. Since then, research and development to improve tubes never stopped.

\bigskip

The traveling-wave tube brought many benefits from the point of view of electrical laboratories. Performance showed the device to be more suitable for telecommunication usages than other vacuum tubes, at a period where powerful means of message transmission were needed (see next section).
And, for academic purpose, all the theoretical study was to do.
But its main advantage was that, as a new device, each competitors were on equal terms, even small ones. It was for those reasons the French public laboratory CNET went to investigate the TWT immediately after its annoucement \cite{Att96}. But the competition was tough. Indeed, in 1947, CSF obtained the equivalent of 600~000 euros (taking into account inflation) by the French Government for a market research on the new electrical equipment after announcing having made a 2.7~GHz TWT working at 200~mW.

In the summer of 1947, Kompfner attended to the 5\textsuperscript{th} IRE Electron Tube Conference at Syracuse University, New York. At this conference, he discovered TWT had become an important subject, both theoretical and experimental, of many laboratories and industries all over the world. 
He was impressed by advancements achieved, but also worried \cite{kom76} that the researches in the United Kingdom on these subjects had not progressed as much. 
After returning to his country, Kompfner with British Admiralty representatives aimed to report on this conference and to plead for a special effort by Britain to regain the initiative. 
They met the UK Coordination of Valve Development Committee (CVD) composed by government and industrial tube representatives and managed by D.~C.~Rogers of STC\footnote{The Standard Telephones and Cables Ltd.\ (STC) was the British division of the Western Electric.}. Aware of the benefits for communication of TWTs compared to other amplifiers, one of their research teams led by Rogers achieved \cite{rog49,rog53} an efficient 4~GHz TWTA, which will later be used for commercial applications. Under Pierce's recommendation, Kompfner joined the Bell Labs in 1951. 

\bigskip

In parallel, the active TWT developments led to the discovery of the backward-wave oscillator (BWO) (a.k.a.\ carcinotron\footnote{CSF, now TED, still uses the trade-name carcinotron because of the crayfish which swims backwards (in Greek ``karkinos'') \cite{gue52}.}). The BWO is basically a TWT where the wave is propagated in the opposite direction to electrons. Its main use is for military scramblers. Invention of BWO was simultaneously and independently presented by Kompfner from Bell Labs, and by Bernard Epsztein from CSF, at the IRE Electron Tube Conference in Ottawa, Canada, on June 1952 \cite{gue52,kom76}. 
It also seems that a Soviet team secretly invented the BWO back in 1948 \cite{pch03}.
In the same vein, consecutive progress to improve klystrons, in particular to expand their bandwidth, led to the invention of the extended interaction klystron (EIK) \cite{wes57,cho61}, a device ---similar to the rugged coupled cavity and the interdigital line TWTs--- which tries to combine the advantages of both klystrons (ruggedness, and high power capability) and TWTs (larger bandwidth).

\bigskip

The major advances in the United Kingdom, France, the United States, Japan and the Soviet Union brought those countries to the forefront of traveling-wave tubes' manufacturing, and to their commercial applications.

\section*{Wireless radio-relays after 1945}

After the Second World War, European countries needed to rebuild their economic power. While the priority was first given to transport grids, power generation and distribution, the transmission networks progressively gained an interest, notably under the pressure of the North Atlantic Treaty Organization (NATO), for tactical communication purposes, and because of the public's interest for a new medium shelved during the war: the television.
TV emissions were retransmitted in some big cities since the 1930s, and only for a small audience, so the number of programs and channels was limited. Those transmissions were done by radio-wave from low range stations directly to the customer. But to provide TV over countries, it was needed to send the programs to each stations, and there were two available means of propagation: coaxial cables and wireless signals.

The cable option was the oldest one and has been practiced sufficiently long to be firmly established. It was the option chosen, in 1951, for the Birmingham--Manchester cable  \cite{hal52}. 
The Hertzian-wave approach was new\footnote{The world's first experimental microwave (1.7~GHz) radio link was demonstrated in 1931 between Calais and Dover (40~km long) \cite{koh31}.} to broadcast television on country-scale but was promising, and was the option taken the same year for the London--Birmingham radio-relay \cite{Cla51} established with triodes. Because of the Earth curvature, this choice required  relays 50--70~km apart.

\bigskip

In 1948, the British Television Advisory Committee recommended the extension of television diffusion to cover 80\% of the United Kingdom.
Hence, to provide TV in Scotland, the British Broadcasting Corporation (BBC) and the Post Office considered the best means of propagation was the use of wireless waves instead of cables, and they signed a contract with STC to establish the first commercial microwave radio-relay systems in the world using traveling-wave tubes \cite{daw54}. The system needed to carry television signals between Manchester and the Kirk o'Shotts transmitting station, near Edinburgh, at more than 350~km, using seven intermediate relays.
The line was activated\footnote{In parallel with TWTs of STC, the microwave radio-relays between Manchester and Edinburgh used also 2~GHz triode valve amplifiers from the General Electric Company \cite{Bra95}, but this system was later ousted by the one with TWTs.} in August 1952, and was the first application of the, so called, super high frequency (over 3~GHz) outside North America \cite{unk51,Bra95}. With this achievement, the traveling-wave tube demonstrated having improved performances for television signals compared with triodes and klystrons. From $45\,000$ television receivers in 1948, the country went to more than 2~million ones five years later, and the 80\% goal was reached in 1957.

Since at this time, the United Kingdom had been equipped with both cables and radio-relays, an early economic comparison between them was performed \cite{fau52} with the conclusion that, for almost the same service, the cost of the cable system was inevitably higher and required many more repeater stations than a radio-link. But cable required less maintenance charges, and unnecessary expenditure could be avoided when sharing facilities with pre-existing telephone cables. However, microwave radio-relay systems were young and steadily improved.

\bigskip

On the other side of the Channel, France possessed a large telephonic network composed of coaxial cables connecting big cities. Those lines were able to transmit several hundreds of phone calls, but just enough to send one TV program, and pulling additional cable for each wanted signal was an expensive option. Also, a more serious problem was the fact that those cables were designed for multichannel telephony and not well suited for TV transmission.
The Postes, T{\'e}l{\'e}graphes et T{\'e}l{\'e}phones (PTT) administration decided to use the microwave option and they started a collaboration with CSF to upgrade the network of wireless transmission. 
First tests were completed in July 1951, by making the connection between Paris and a radio tower in Bois de Molle, Beauvais, 60~km away. 
The emitter provided by CSF comprised a broad bandwidth klystron to obtain a linear frequency modulation with an output at 1~mW, then the signal was amplified by a TWT up to 1~W for a base frequency at 4~GHz (similar to STC tubes). In 1953, they finished the wireless liaison between Paris and Lille, distant of 230~km. This liaison\footnote{The first French wireless radio-relay of 1~GHz with telephone commercial use was built in 1951 by the Compagnie Fran\c{c}aise Thomson-Houston (CFTH) for the Radiodiffusion - T{\'e}l{\'e}vision Fran\c{c}aise, from the Eiffel Tower, Paris, to the city hall belfry of Lille, with two relays \cite{ang52,mar52}. This line did not involve TWTs and was replaced after 1953 by the other one presented.} was the first in the world to provide both telephone and television using the same transmitters \cite{For51,mar51,gut52}. 

\begin{table}[ht!]
\centering
\caption{Commercial helix traveling-wave tubes used in early microwave radio-relays in the United Kingdom, France and the United States. The CV2188 was the first commercial tube and was used in the Manchester--Edinburgh line. The TPO~921 is the direct sucessor of the TPO~851 used in the Paris--Lille line. The type 7812 was a TWT (third stage) used on the Tokyo--Osaka line. The 444A (developed by Bell Labs) was used in the TH system covering the United States. For Color-Television (CTV), a much wider bandwidth is required than for black and white. Sources: \cite{rog53,vog57,saw56,mcd60}.}
\label{tab:2}       % Give a unique label
\begin{tabular}{lcccc}
\hline\noalign{\smallskip}
Manufacturer & STC & CSF & Shibaura & Western Electric \\
Name & CV2188 & TPO 921 & 7812 & 444A  \\
\hline\noalign{\smallskip}
Year & 1952 & after 1953 & 1954 & 1960 \\
Capacity & 1 TV &  1 TV + 240 ph. &1 TV  &1 CTV + 420 ph.  \\
$\quad \quad \; \; \;$ or &  & 720 phones & Hundreds & 1860 phones \\
Frequency & 3.6-4.4 GHz & 3.8-4.2 GHz & 3.5-4.3 GHz & 5.9-6.4 GHz\\
Power Out & 2 W & $>$ 2 W & 3.5 W & 5 W \\
Gain & 25 dB & 28-30 dB & 17 dB & 30 dB\\
Efficiency & $\approx$ 1\% & 20\% & ? & 23\% \\ %Mass & $>$ 5000 g & ? & ? \\
Life Time & 3 500 hours & thousands hours & ? & 10 000 hours\\
\noalign{\smallskip}\hline
\end{tabular}
\end{table}

Before June 1953, the Paris--Lille liaison was extended to London and Brussels with CSF hardwares. This line and all other TV networks available in Western Europe ---including cable lines--- were used for the live transmission of the coronation of Queen Elisabeth~II, on June~2. Broadcast\footnote{The live multinational TV transmission of the coronation of Queen Elisabeth~II in 1953 was organized by the European Broadcasting Union. This success led one year later to the creation of Eurovision.} simultaneously in at least six countries to several million of TV spectators, it was the world's first major TV event \cite{smi53}. Millions of other people were able to see the coronation after tapes were sent\footnote{To provide videos of the coronation to the rest of the world, and especially the Commonwealth, tapes were sent by airplane. This was a common practice until transoceanic communications by satellites were implemented.} around the world.

\bigskip

On April 1954, Japan showed its advances when opening the Hertzian-wave line Tokyo--Nagoya--Osaka, more than 460~km long. This line was established for the Broadcasting Corporation of Japan (NHK) to start regular TV emissions, and also for the Nippon Telegraph and Telephone Public Corporation (NTT) to rebuild the country's telephone service which had been wiped out (nearly 80\% of the service) by bombing.
It was the first microwave radio-relay system of the country and the system was equipped with three-stage TWTs working at 4~GHz and built by the Tokyo Shibaura Electric Co ---nowadays Toshiba Corporation--- \cite{nom54,saw56,you65}. Ten years later, in 1964, NHK covered 81\% of the Japanese population for 13~millions viewers, and NTT had 8.6~million telephones in service.

\bigskip

During this time, the pioneers of microwave radio-relay systems were the United States. Immediatley after the war, the urgent need to provide television throughout the country pushed AT{\&}T to develop their microwave radio-relay systems. But it was before first commercial TWTs, so Bell Labs began a live-demonstration by building the New York--Boston wireless line \cite{dur47,tha49}, which was 350~km long (with eight relays), using 4~GHz triodes \cite{mor49}. This system (called TD-2) could handle 240 (then 600) telephone channels at once and was used afterwards by AT{\&}T for commercial applications. In 1951, the New York--San Francisco line was established\footnote{The line was $4\,700$ km long with 117 stations about 52~km apart and cost $380$ million dollars (with inflation).} \cite{roe51} and the system was extended to more than $70\,000$~km with relays by the end of 1960. 
From $155\,000$ television receivers in 1948, the country went to more than 15~million ones four years later.
There are several reasons why triodes were privileged in the first instance in the United States. The main one is historical. In 1912, AT{\&}T had bought de Forest's triode patent, and they already used it as a repeater for the 1914 New York--San Francisco telephone cable, the world first transcontinental telephone line. Four decades of improvements in all its aspects made the triode hard to dethrone.
Also early microwave tests (1945) were done before the TWTs was known.
But at the end of the 1950s, the triodes were replaced by the 6~GHz TWT newly developed by Bell Labs, carrying at least 1860 telephone channels (called TH system) \cite{mcd60,jar64}.
 %Also, in the United States, early microwave tests were done before the TWTs was invented, while in the rest of the world fell behind probably because of the World War II aftermath.

\bigskip

Only 10~years after its (third) invention in 1942, the traveling-wave tube was already commercially used (cf.\ table~\ref{tab:2}), which is an exceptionally short time since the device was not really optimized at this time, and some companies had sold TWTs after only 5~years of development. Several kinds of vacuum electron tubes, like triodes, were used, but the appearance of traveling-wave tube amplifiers (TWTAs) provided large bandwidth signals ideal for long-range TV-transmission, and replaced other vacuum tubes. 
It was not the first vacuum tubes used for long-range communication, but it surely boosted the development of microwave radio-relay systems for television around the world between the 1950s and the 1980s. At the same time, international agreements on favored standards (including frequency band designation, see table~\ref{tab:1}) were achieved by organizations from a wide range of countries including the Soviet block. 

\bigskip

The next step was sending those relays in outer space.

\section*{Extra-terrestrial radio-relays}

It is hard to say with certainty who was the first person to write about communications in space. 
Probably inspired by the 1865 novel \textit{From the Earth to the Moon} by Jules Verne \cite{ver65}, and of Percival Lowell's books \cite{low96,low06} dealing with Martian life, the Soviet science-fiction writer Aleksey N.~Tolstoy (a remote relative of Leo) wrote \textit{Aelita} in 1923, a novel about explorers leaving the Earth with a rocket and finding Atlantean descendants on Mars \cite{tol85}. The novel was adapted, a year later, in a silent film with the same title.
Not written in the novel, the film's narrative thread is the moment when every radio stations on Earth receive the same message from Mars: ``Anta... Odeli... Uta''. %The movie was acclaimed by the Soviet press, and became a great success in the Soviet Union. 

On the other side of the world, in the United States, the science-fiction writer Hugo Gernsback published a serious seminal article titled \textit{Can we radio the planets?} \cite{ger27}. In this article, he wrongly assumed that if the Earth, meteors and stars are composed predominantly of iron, then the Moon should be too, and according to Hertz works ``it would therefore make an excellent reflecting medium''. So, he proposed to use the Moon as a reflector to determine the existence of the Kennelly-Heaviside layer and he dealt with interplanetary emission. This was followed, two years later, by a study from a geophysicist on commercial Hertzian-wave propagation on the Mars ionosphere \cite{hul29}. 

\bigskip

Inspired by Gernsback's work, a few papers \cite{mof46,gie48} proved the feasibility of using the Moon as passive reflector, like a mirror, to achieve transcontinental communication, leading later to the first signals sent and received through space \cite{goo46}. In the early 1950s, the U.S.\ Naval Research Laboratory (NRL) developed the Communication Moon Relay project (a.k.a.\ Operation Moon Bounce) \cite{But97} and they transmitted the first vocal message in space sent back to Earth, on July~24, 1954. At this time, ground stations used 10~kW klystron amplifiers. This project was used in 1959 by the U.S.\ Navy, in a context of cold war, to create a backup line between Washington, D.C.\ and Hawaii or U.S.\ fleets. Indeed, ionospheric storms could cut off radio transmissions. The system\footnote{The Communication Moon Relay project was originally from a classified program of espionage, the Passive Moon Relay (PAMOR), to monitor Soviet communications reflected on our natural satellite \cite{But97}.} was made public in 1960. 
But this concept was difficult to implement because the distance and the weak fraction of radiation reflected off the Moon, impose powerful devices, making TV transmissions difficult. In addition, there is a 2.5~seconds delay to send any signal ---too much for telephone conversations--- not to mention the fact that the Moon is visible only up to twelve hours a day.
In conclusion, our natural satellite was not a perfect relay but it gave the idea to use much closer reflectors.

 %\footnote{Gernsback created one of the first science fiction magazines called \textit{Amazing Stories} in 1928.} 
 
 \bigskip

%Arthur C.~Clarke, a British scientist and science-fiction writer, is undoubtedly a pioneer responsible for this fever.
Twelve years before the launch of the satellite \textit{Sputnik~1}, the British scientist and science-fiction writer Sir Arthur C.~Clarke\footnote{Sir Arthur Charles Clarke's vision of the space conquest was inspired by traumatic \textit{V-2} rockets developed during World War II. Before writing on space stations for telecommunications, he wrote another article \cite{Cla45a} in \textit{Wireless World} to propose a peaceful use of those \textit{V-2} at geosynchronous orbit (GSO) for research on the ionosphere. Indeed, the space race started immediately after the war, when the United States and the Soviet Union collected as many \textit{V-2} rockets as possible and captured German scientists.} wrote two papers ;
the first one was private \cite{Cla45b} and only given to his colleagues at the British Interplanetary Society, while the other was published \cite{Cla45c} in the journal \textit{Wireless World}, titled ``Extra-Terrestrial Relays'', and both dealt with the usefulness of putting satellites at a geosynchronous orbit (GSO) for communications. GSO (a.k.a.\ Clarke orbit) was imagined by the Russian theorist Konstantin E.~Tsiolkovsky\footnote{Konstantin Eduardovich Tsiolkovsky is considered as one of the fathers of modern astronautics for his theoretical developments, and was a science-fiction writer. In addition with geosynchronous orbit, he is also granted for the rocket equation and for the multistage rocket concept, a key element in the success of \textit{Sputnik~1}.}, who proved that an object orbiting at $36\,000$~km above the equator would appear as stationary from the Earth, because its orbital period is 24~hours. 
For comparison, at 400~km, objects have a period of an hour and a half, while the Moon, at $362\,000$~km, has a period of 28~days.
Clarke's revolutionary idea was to place on this orbit three space stations covering all the planet, intended for censorship-free global TV and radio-telephones. Since those objects stay in the sky at the same point, it is easy to point an antenna at them. In Clarke's proposals, space stations worked via some ``solar engines'' and were regularly supplied from Earth. Clarke estimated that GSO could be reached by artificial crafts in perhaps half a century ahead.
While the publication of his article did not collect much audience in 1945, the appearance of the first satellites made Clarke famous\footnote{His article \cite{Cla45c} became so popular that it was reprinted in the centenary issue of the \textit{Electronics World} \cite{jos13} (formerly \textit{Wireless World}) as one of the most influent paper of the journal.} as a pioneer in the domain.

The next year, in May 1946, the members of the Project RAND\footnote{``Research ANd Development'', now the RAND Corporation, is an American think tank founded by the U.S.\ Army Air Force and the Douglas Aircraft Company.}, working for the U.S.\ Army Air Force, started to investigate Clarke's ideas. They wrote a complete report \cite{pro46,dav88} dealing with GSO satellites as relays for communications, but also included aspects on military applications, scientific research, weather reconnaissance, interplanetary travels, and practical analysis, like orbit trajectories and payload available. They estimated that this achievement would bear considerable repercussions to the world, comparable to the Wright brothers' success, or the explosion of the atomic bomb. 
But the initiative to build the first communication spacecrafts came from civilian commercial investigations.

\bigskip

Meanwhile, Pierce was also known ---under the pseudonym J.~J.~Coupling--- as a science-fiction writer and a precursor in the field of space communications. He wrote a novel mentioning the possibility of reflecting radio-waves from the Moon and interplanetary radio signals \cite{pie52}. 
Just after, he started to seriously work on his ideas with his colleagues at Bell Labs. In 1955, he proposed \cite{pie55} the first paper addressing the economic viability for orbital radio-relays. 
At this time, AT\&T and the British Post Office were building the first transatlantic telecommunications cable (TAT-1) \cite{sch08}, providing 36~telephone channels all at once, where existing cables were only for telegraph. TAT-1 was inaugurated in 1956 and costed approximatively 42~million dollars. 
Pierce analysed that to provide one television signal ---or 1080~telephone channels--- across the ocean, you would need a billion dollars or more to put additional cables. 
If building a spacecraft is worth this billion, then the concept appeared immediately more suitable. 
Pierce estimated that relays in space would not compete with microwave radio-relays over land, but would certainly be feasible for transoceanic communication. 
In his proposal, satellites are not at the geosynchronous orbit (GSO), but instead, are at a lower height called Low Earth Orbit (LEO), imposing for communications to wait until satellites appear overhead emitters and receivers. To remedy the lack of permanent link with the ground, the system needs several other identical satellites ---a configuration called a satellite constellation--- to ensure that at least one satellite be visible at any time.
LEO crafts can be sent with smaller rockets or carry heavier payloads. 
Pierce also mentioned the possibility to use either passive crafts to reflect signals, like with simple spherical mirrors, or active crafts where signals are re-amplified before being sent back.

\bigskip

The idea for extra-terrestrial relays was established and only needed a remarkable instigation to appear.

\section*{First communication satellites}

On October~4, 1957, the Soviet Union reached the outer space with the first man-made object orbiting\footnote{The \textit{V-2} rockets, used in the 1940s by German, then after the war by American, British, and Soviet, were probably the first man-made objects in space but they were not orbiting. They also took the first Earth pictures and videos \cite{whi52}.} around Earth, the satellite \textit{Sputnik~1}, launched from Tyuratam. Because it was orbiting at very low height, a repetitive steady beep, emitted from a 20--40~MHz pentode, was receivable by any radio amateurs throughout the world. There is no doubt the Soviet achievement was the disruptive element that started the space race once for all in the United States.
It led to the first American satellite \textit{Explorer~1} launched in February 1958 from Cape Canaveral, Florida. Less than three years later, more than a hundred objects had been launched\footnote{For instance, American sent \textit{SCORE} (Signal Communications by Orbiting Relay Equipment) in December 1958. The satellite carried a recorded voice message from President Eisenhower, continuously emitted like the \textit{Sputnik} beep. But the device was not a relay for live telecommunication. In April 1960, the American \textit{TIROS-1} (for Television Infrared Observation Satellite) became the world's first weather satellite. The craft was composed of two cameras sending down videos of the Earth \cite{log95}.} into space, successfully or not, by Soviets and Americans.

\bigskip

The \textit{Echo} project, started in 1956 by the National Advisory Committee for Aeronautics (NACA), was originally a mission to measure the density of the upper atmosphere by observing a 3.5~metres diameter balloon-like satellite. During the U.S.\ Sputnik crisis in 1958, NACA was dissolved to become the National Aeronautics and Space Administration (NASA). At this time, Pierce and Kompfner ---after he joined Bell Labs in 1951--- realized that it would be possible to use spherical mirrors as passive reflectors to test space communications. NASA accepted this suggestion, and the \textit{Echo} project became NASA's first communications satellite project. On August~12, 1960, they launched a plastic sphere, with aluminized surface, named \textit{Echo~1}, large of 30~metres in diameter. It became the first artificial satellite that actually relayed a real-time voice message, from Holmdel\footnote{The Bell Labs horn antenna of Holmdel, built in 1959 to support the \textit{Echo} project, is also famous because of its association with the discovery of the cosmic microwave background, the oldest light in the universe, by Bell Labs employees A.~Penzias and R.~Wilson who were awarded the Nobel Prize in 1978.}, New Jersey to Nan\c{c}ay (near Bourges), France. It was a passive object because there were no electronic systems to amplify the signal aboard. 
The success of \textit{Echo~1} proved it was possible to send a message through space via man-made relays. 
However, the limited communication capacity of a single voice channel highlighted that passive reflectors would not have a lot of applications. 
A similar project between the same stations was also being conducted using the Moon as a passive reflector. 
Those two projects were crucial to the improvement of tracking techniques ---\textit{Echo~1} was visible and usable as relay only 5~minutes per passage over the Atlantic--- and ground station equipments. 
Also, while ground stations at this time used klystrons as power amplifiers, these were quickly replaced by TWTs for the \textit{Telstar} project.

Thereafter, to demonstrate the reliability of the active satellite, and to keep their advantage in long distance communications, AT\&T and Bell Labs approved the \textit{Telstar} project initiated by Pierce, Kompfner, and some or their colleagues at Bell \cite{cra63}. Like Clarke's idea, an active relay would amplify signals before sending them back to Earth. But reaching the geosynchronous orbit seemed too difficult for the researchers, instead \textit{Telstar} satellites were planned for a low altitude orbit (LEO), like \textit{Echo~1}, so they could only be seen, at the same moment, by two ground stations for a maximum of twenty minutes across the Atlantic, with an orbital period of 2.5~hours around our planet. 
The original idea was to make fifty five satellites (a constellation) to cover almost the entire Earth surface, linked with twenty five ground stations at any time.
 Pierce estimated an expense of 500~million dollars.
\textit{Telstar} satellites were nearly spherical polyhedra of 88~centimetres diameter for 77~kilograms. They were composed of the new technology required for satellite communication, like thousands of transistors for other various electric systems, solar cells for power generation, and a 4~GHz helix traveling-wave tube amplifier from Bell Labs \cite{bod63}. In fact, all the satellite's active elements were solid-state devices (transistors) excepted for the TWT amplifier.
On the U.S.\ East coast, Bell Labs built a ground station at Andover, Maine. 
On the European side, a French--British rivalry led to the construction of two ground stations, one at Goonhilly Downs, Cornwall, by the British Post Office, and one at Pleumeur-Bodou\footnote{The antenna of Pleumeur-Bodou, France, used for \textit{Telstar~1} signals was completed on July~7, 1962, viz.\ three days before the launch of the spacecraft. }, Brittany, by the French PTT. All three ground stations exploited a 2~kW TWT with coupled cavities as slow-wave structure \cite{col63}.
\textit{Telstar~1}, launched on July~10, 1962, was the world's first active satellite for telecommunication, and transmitting\footnote{The first transmission by \textit{Telstar~1} on July~10, 1962, is a telephone conversation between the chairman of AT{\&}T and Vice President Johnson, from Andover to Washington. This was followed by TV emission from the United States to France. France sent back footage of Yves Montand interpreting ``la Chansonnette'', while across the Channel, Britain sent back a color test card (the world's first transatlantic color transmission). Then America and Europe sent each other various videos and messages, like footages of Mount Rushmore, the Statue of Liberty, the Eiffel Tower, or a base-ball match and a declaration of President John F. Kennedy, broadcast to the public by local television organizations.} television across the Atlantic Ocean. 
In her following Christmas message, Queen Elizabeth II mentioned that ``this tiny satellite has become the invisible focus of a million eyes'' \cite{tit13}.
After this success, scientific, engineering, financial and political forces moved forward in developments for satellite communications which also increased the interest of TWTs.

\bigskip

Just five months later, on December 14, 1962, NASA launched \textit{Relay~1}, a similar satellite with the same features as \textit{Telstar~1} but built\footnote{When NASA initiated the \textit{Relay} project, AT\&T and Hughes Aircraft Company tried to win the contract to build the two spacecrafts but finally NASA selected RCA.} by RCA and with a longer lifetime in orbit. \textit{Relay~1} was the first satellite to broadcast between the United States and Japan. The spacecraft is also known for its large number of recorded anomalies, including the 60~minutes time to warm up the traveling-wave tube (usually it was only taking around 3~minutes) \cite{But97}.

\bigskip

On another hand, Haeff ---after he left RCA to briefly join the U.S.\ Naval Research Laboratory (NRL)--- joined the Hughes Research Laboratories (HRL) of the Hughes Aircraft Company\footnote{The Electron Dynamics Division, of the Hughes Research Laboratories (HRL) of the Hughes Aircraft Company, became L3 Electron Devices in 2005, and still is a major space TWT manufacturer.} in 1950, and became rapidly vice president and director of research\footnote{Haeff left Hughes in 1961 and his position was given to Field.} in the company. Hughes was developing their own TWTs, and it was more and more interested in the space conquest.

When Pierce and Kompfner were conceptualizing \textit{Echo} and \textit{Telstar} projects, they published, in 1959, an article \cite{pie59} in the \textit{Proceedings of the IRE} giving their view of transoceanic communication. For them, a constellation of at least twenty four active satellites orbiting at LEO height would be enough to cover the world. This vision was not shared by Harold A.~Rosen, an electrical researcher at Hughes. Unaware of Clarke's work, for him the future of satellite communication was achievable by reaching the geosynchronous orbit (GSO). Agreeing with him, Haeff formed a task force, led by Rosen, to initiate the \textit{Syncom} design. Unlike low orbit, to reach GSO, one must
 drastically reduce the weight of the launched object. A part of the \textit{Syncom} success was due to the lightweight traveling-wave tube built by John~T.~Mendel\footnote{After earning his Ph.D.\ from Stanford University in 1952, John Thomas Mendel was employed at Bell Labs. Then he joined Hughes until he became a vice president of the company.} at Hughes, with a weight of half a kilogram \cite{hig62}. The total mass of the spacecraft was less than 40~kg, compared to the 170~kg of \textit{Telstar~1}.
Following an enthusiastic report on the feasibility of the project, Hughes Aircraft Company funded its construction, and after a previous failure, \textit{Syncom~2} became on July~26, 1963, the first geosynchronous satellite; it was equipped with Hughes' TWTs. The major advantage of a geosynchronous satellite is that ground station can keep the link with the satellite at any time, easing the tracking. In 1964, \textit{Syncom~3} orbited over a sustained period of time, and transmitted the Summer Olympics from Tokyo to the United States.

\bigskip

Understanding the huge impact of those new orbital radio-relays, the U.S.\ government funded, in August 1962, the Communications Satellite Corporation (Comsat), a government-owned telecommunication company recognized by western countries. Its first satellite \textit{Intelsat~1} (a.k.a.\ \textit{Early Bird}) was the world's first commercial communications satellite. Built by the Hughes Aircraft Company, it was launched in April 1965 on a GSO. 
The next step for those satellites was to increase their area capability, with more powerful amplifiers with a wider bandwidth. The second series of Intelsat crafts integrate a multiple-access capability by carrying four 6~W traveling-wave tubes for each spacecraft that could operate simultaneously.
But when Comsat was funded, President Kennedy gave it the monopoly on space transmissions \cite{But97}. AT{\&}T's \textit{Telstar} project\footnote{Bell Labs designed and built six \textit{Telstar} spacecrafts, but only two were launched. NASA negotiated an excellent deal with AT\&T because NASA's contribution to the project was limited to launch services, but they claimed the project to be supported by them, and they even published results of the experiment as a NASA publication, while it was originally issued as articles in the Bell Telephone technical journal.} immediately vanished because Comsat only bought satellites from the Hughes Aircraft Company, preferring the GSO configuration.

\begin{table}[ht!]
\centering
\caption{Example of helix space traveling-wave tubes used in early satellites. The M4041 was the first TWT in space. \textit{Applications Technology Satellites} (\textit{ATS}) were NASA's experimental probes based on \textit{Syncom} design. Sources: \cite{fel65,unk63,unk63b,unk65,kor01}.}
\label{tab:3}       % Give a unique label
\begin{tabular}{lcccc}
\hline\noalign{\smallskip}
Manufacturer & Bell Labs & RCA & Hugues & AEG-Telefunken  \\
Name & M4041 & A-1245 & 384H &  TL 4003 \\
Satellite & \textit{Telstar 1} & \textit{Relay 1} & \textit{ATS 1} & \textit{Symphonie~A}  \\
\hline\noalign{\smallskip}
Launched & 1962 & 1962 & 1966 & 1973 \\
Frequency & 3.7-4.2 GHz & 4.05-4.25 GHz & 3.96-4.12 GHz  & 3.7-4.2 GHz \\
Output Power & 2 W  & 11 W & 4 W  & 13 W\\
Gain & 40 dB & 35 dB & 36-40 dB & 46 dB \\
Efficiency & $<$ 10\%  & 12\% & ? & 34\% \\
Mass & $>$ 1000 g & $<$ 2000 g & $<$ 567 g &  640 g \\
Life Time & 100 000 hours & $>$ 5 years & 50 000 hours & ?  \\
\noalign{\smallskip}\hline
\end{tabular}
\end{table}

In the Soviet block, telecom satellites started with \textit{Molniya 1-1}, launched less than three weeks after \textit{Early Bird}, in 1965. 
 %with the third generation of Molniya satellite equipped with three 30 watts TWTAs. 
This satellite and its successors were put on elliptical orbits called Molniya orbits ---one category of Highly Elliptical Earth Orbit (HEO)--- with a center largely shifted from Earth, enabling them to appear over northern latitudes most of the day.

\begin{table}[h]
\centering
\caption{Microwave bands of frequencies (also referred as hyper-frequencies, or centimetre and millimetre waves) from IEEE standards \cite{ieee} established in 1976. Higher frequency enables larger flow of information, but are more subject to atmospheric attenuation. Present satellites operate from L to Ka-Band.}
\label{tab:1}       % Give a unique label
\begin{tabular}{r|cccccc}
\hline\noalign{\smallskip}
Band & L-Band & S-Band & C-Band & X-Band & Ku-Band  \\
GHz & 1 to 2 & 2 to 4 & 4 to 8 & 8 to 12 & 12 to 18 \\
& \\
Band & K-Band & Ka-Band & Q-Band & V-Band & W-Band  \\
GHz & 18 to 26.5 & 26.5 to 40 & 33 to 50 & 40 to 75 & 75 to 110 \\
\noalign{\smallskip}\hline
\end{tabular}
\end{table}

\bigskip

It is not a coincidence that the main U.S.\ manufacturers of traveling-wave tubes (Bell Labs, RCA, and Hughes) at this time, were also those who supplied space communication devices. Researchers who developed TWTs were conscious this device ---reliable, effective and light--- was perfectly adapted for the space conquest\footnote{To reach the Moon, \textit{Apollo}'s Command Modules (CSM) were equipped with S-Band TWTAs \cite{bal68,ros72}.} (cf.\ table~\ref{tab:3}).
Progressively, transcontinental telephone and television communications were enabled everywhere, including specialized satellites, like for maritime ships or airplanes (\textit{Inmarsat}), and Direct-to-Home TV broadcasting (\textit{Intelsat}, \textit{Eutelsat}, \textit{Galaxy}, \textit{Astra}), and new services, like satellite phones (\textit{Iridium}, \textit{Globalstar}), GPS (cited below), Internet access by satellite (\textit{Wildblue}, \textit{KA-SAT}), or numerical radio diffusion (\textit{XM Radio}, \textit{Sirius}).
They also offered high-definition images of the Earth, including contributions in weather science, and images of solar system objects.
In the same time, Rosen\footnote{Harold Allen Rosen won, in 1995, with Pierce, the Charles Stark Draper Prize for ``development of communication satellite technology''. He was involved with the majority of Comsat crafts built after the 1960s for Hughes Aircraft Company, and then the Boeing Company.}, Clarke\footnote{Writer of almost 100~books, Clarke is one of the most influential science-fiction authors of his time with Robert Heinlein and Isaac Asimov. The world remembers him mainly as the writer of \textit{2001: A Space Odyssey}, simultaneously published and released as a movie by Stanley Kubrick in 1968.}, and Pierce\footnote{In addition to TWTs, satellite communications and science-fictions, his pioneering spirit brought John Robinson Pierce to contribute in information theory with Claude E.~Shannon, and in music theory. With his colleague Max Mathews and others, they released \textit{Music from Mathematics}, an album completely played by an IBM 7090 computer. One of their songs, \textit{Daisy Bell} (a.k.a.\ \textit{Bicycle Built for Two}), was later interpreted by the fictitious artificial intelligence HAL 9000, in Stanley Kubrick's \textit{2001: A Space Odyssey}. He is also the neologist of the term ``transistor''. Beside computing science, transistors can be used as solid-state power amplifiers (SSPAs), and ironically are the main competitors of vacuum electron tubes.} started to share the title of ``fathers of satellite communications''.

%If the traveling-wave tube have not been invented, human would still have explored space, but it 

\section*{The European space conquest}

Four days after the launch of \textit{Telstar~1} in 1962, a few western European countries joined the space conquest race by signing an agreement to establish two new space agencies: the European Space Research Organisation (ESRO) which would build scientific probes, and the European Launcher Development Organisation (ELDO) which would focus on a launcher.
But those agencies fell behind for many reasons, like the number of member states with different space policies and budgets, or the issue of users. Indeed, since western Europe is relatively small and not crossed by an ocean, a large telecommunication program could be considered as a superfluous luxury compared to terrestrial options.
Meanwhile, some European countries progressed with theirs own finances\footnote{The United Kingdom operated \textit{Ariel~1} (a.k.a.~\textit{UK-1}) in 1962, a satellite built by NASA. Then, the Italian Commissione per le Ricerche Spaziali (CRS) sent its own spacecraft \textit{San Marco~1} in 1964 with a U.S.\ rocket. Finally, the Centre National d'{\'E}tudes Spatiales (CNES) ---founded in 1961 by President de Gaulle--- launched \textit{A-1} (a.k.a.~\textit{Ast{\'e}rix}) with a French launcher \textit{Diamant}, in 1965. Those three spacecrafts carried no TWTs.}.
Until 1975, ESRO built eight scientific satellites ---one failure due to the rocket--- sent by American rockets. Indeed, ELDO had a series of failures in attempting to develop a European launcher. 
This led to the merging of ESRO and ELDO, to create the European Space Agency (ESA) in 1975. Since then, they continued the quest for a launcher with the Ariane project. 

\bigskip

During the 1960s, President de Gaulle and Chancellor Adenauer had acted for French-German cooperation. To balance the communication monopoly of the two superpowers, a consortium was created between French and German organisations; \textit{Symphonie~A} was launched on December 19, 1974, with the American rocket \textit{Delta}, from Cape Canaveral, Florida, and became the first European satellite for telecommunication. The 13~W TWTAs ---the first space TWTs built in Europe\footnote{Before that, there were other European vacuum electron tubes in space, like the 10~W output, S-Band triodes from Siemens ---escorting Hughes' TWTs--- in the NASA's \textit{Mariner} program \cite{fel65,kos82}. \textit{Mariner~2} is the first probe to achieve a planetary flyby (Venus), followed by \textit{Mariner~4} (Mars) and \textit{Mariner~10} (Mercury) \cite{sid02}.}--- working in C-Band (see table~\ref{tab:1}) were provided by the German AEG-Telefunken company.
Just after, the newly merged ESA financed a series of two experimental telecommunication satellites called \textit{Orbital Test Satellite} (\textit{OTS}), and equipped with Ku-Band TWTAs from the French Thomson-CSF. After a failure of the rocket, \textit{OTS-2} was put in GSO in 1978.
The consequence of this demonstration was the creation in 1977 of the intergovernmental European Telecommunications Satellite Organization (Eutelsat) to develop space communication in Europe, financed at 60\% by ESA \cite{But97}.
Eutelsat started operations using in first instance \textit{OTS} crafts with ESA consent, then brought its own satellites. 
The \textit{OTS} program served as the forerunner of the \textit{European Communication Satellite} (\textit{ESC}), with four satellites launched in the 80s, including \textit{ESC-1} launched with the new European \textit{Ariane 1} launcher.

In the continuation of \textit{Symphonie~A}, another French-German consortium was created to develop TV broadcasting in the two countries. Between 1987 and 1990, they successfully launched four satellites: \textit{TDF 1} and \textit{2} for France, and \textit{TV-SAT 1} and \textit{2} for Germany.
Each was sent into GSO using \textit{Ariane 2} and \textit{4} launchers. \textit{TDF 1} was equipped with 240~W Ku-Band TWTAs from Thomson-CSF, the most powerful tube at this time.
In comparison, currently state-of-the-art Ku-Band TWTAs work at 200~W maximum. Operators do not need more power, mainly thanks to improvements in antennas.

\bigskip

Meanwhile, Japan had also joined the party as another foreign competitor for the United States and Europe. In February 1983, the National Space Development Agency of Japan (NASDA) launched \textit{Sakura 2a} (a.k.a.~\textit{CS-2a}), the first commercial (and civilian) satellite equipped with Ka-Band TWTAs, using a Japanese \textit{N-II} rocket.

\section*{Present space uses of traveling-wave tubes}

In 2017, we celebrated the sixtieth anniversary of \textit{Sputnik 1} and the beginning of the space conquest. The traveling-wave tube is still widely used in satellites and scientific spacecrafts, even though solid-state power amplifiers (SSPAs) have gained importance and could compete with them nowaday, especially for amplifiers below and up to C-Band. 
Nowadays, the two dominant manufacturers on the space TWT market are the French-German Thales Electron Devices (TED) ---formely Thomson-CSF and AEG-Telefunken--- (cf.\ table~\ref{tab:5}) and the American L3 Electron Devices ---formely Hughes. The Nippon Electric Corporation (NEC) was another important player until it stopped their space tubes commercialisation in the end of the 1990s. The Indian CEERI also entered the competition in 2012, as well as the Chinese BVERI.

\begin{table}[ht!]
\centering
\caption{Examples of recent helix space traveling-wave tubes manufactured by Thales Electron Devices, and used in current satellites. The TH 4626, and the TH 4606C will respectively be installed on the \textit{James-Webb Space Telescope} and the probe \textit{BepiColombo}. For each ones, the life time is estimated to more than 15 years. Sources: \cite{dur15,gas14,bar18,Thales}.}
\label{tab:5}       
\begin{tabular}{lcccccc}
\hline\noalign{\smallskip}
Name & TL 4150  & TH 4795 & TH 4816 & TH 4626 & TH 4606C & THL40040CC \\ 
\hline\noalign{\smallskip}
GHz &  3.4-4.2 & 10.7-12.75 & 17.3-20.2 & $\sim$ 26 & 32 & 37.5-42.5 \\
$P_{\rm out}$ & 150 W & 150 W & 160 W & 50 W & 35 W & 40W \\
Gain & 50 dB & $>$ 50 dB & $>$ 50 dB & $>$ 50 dB & $>$ 50 dB & 48 dB \\
Eff. &  73\% & 68\% & 63\% & 55\% & 54\% & 50\%  \\
Mass & 1000 g & 800 g & 900 g & 700 g & 700 g & $<$890g \\
%Life time & $>$15 years & $>$15 years & $>$15 years & $>$15 years & $>$15 years & $>$15 years \\
\noalign{\smallskip}\hline
\end{tabular}
\end{table}

\bigskip

The Global Positioning System (GPS), initiated by the U.S.\ government for military operations, became freely available\footnote{GPS was allowed for civilian air lines in 1983 by president Reagan after a Boeing~747 of the Korean Air Lines was shot down by the Soviet Air Force in the Sea of Japan.
A navigation error conducted the plane to trespass the Soviet prohibited airspace. } for civilian use in 1983. GPS is currently achieved by a constellation of thirty-one satellites orbiting on Medium Earth orbits (MEO) situated between LEO and GSO, and using only solid-state power amplifiers (SSPAs) because it does not need large data transmissions. However, there exist other global navigation satellite systems to balance the American monopoly and which all use TWTs, like the fifteen \textit{Galileo} satellites ---and ten others scheduled---, financed by the European Union and launched by ESA, or the Chinese \textit{BeiDou-2}, the Indian \textit{IRNSS}, and the Japanese \textit{QZSS}. Indeed, in the L-Band, SSPAs and TWTAs have similar RF performances ---but SSPAs are smaller and less expensive.

The technological revolution of the early 21\textsuperscript{st} century is undisputably the Internet. 
If the main traffic of the net is sent though cable or optical fibre, however, communication satellites were early used to connect isolated big city hubs (important nodes), which redistribute the flux to individual homes and offices. One of major issues of the Internet is the global cover of rural areas.
Some programs try to remedy this, like the French THD-Sat Project, financed by the French government and carried out by CNES, in order to develop the products and technologies dedicated to a new generation of satellites for high throughput Internet access. In particular, THD-SAT is supporting the development of two space TWTAs from TED, one 170~W in Ka-Band for direct communication to individual homes and a second one of 40~W in Q-Band \cite{bar18} for the satellite transmission toward anchor stations. This project is a forerunner of space Q-Band applications.

\bigskip

In 2017, we also celebrate the 40 years of the launch of twin probes \textit{Voyager}. \textit{Voyager~1} is performing its journey at more than 21~billion kilometres from Earth ---140~times the Sun--Earth distance, viz.\ more than 4~times the Sun--Neptune distance, the longest telecommunication range ever done--- and is currently the farthest man-made object still in communication with us thanks to its three TWTAs built by Watkins-Johnson\footnote{The Watkins-Johnson Company was founded in 1957 by Dean~A.~Watkins, a former professor of electrical engineering at Stanford University, and Horace~R.~Johnson, a former head of Hughes Aircraft Company's microwave laboratory \cite{gra96}.}: one in the S-Band, and two in the X-Band \cite{unk77}. The signal takes 19~hours to reach us from the probe, and is captured by a worldwide network of tracking facilities belonging to the NASA Deep Space Network. Her sister \textit{Voyager~2} made a detour near ice giants Uranus and Neptune ---the only spacecraft to have ever visited them--- and is only more than 17~billion kilometres from Earth (or 16~light-hours away). Even now, the probes regularly send us data about cosmic rays or sun magnetometry. Between them flies \textit{Pioneer 10} ---sent with two 8~W TWTs working at 2.3~GHz \cite{sie73}--- but NASA lost communication with the probe in January 2003.

\bigskip

The majority of deep space missions were, are and will be equipped with TWTAs. As a non-exhaustive list, we can mention the following explorer probes: 
\begin{itemize}
\item \textit{Giotto} sent in 1985 ---terminated in 1992--- by ESA to fly by the Halley comet; 
\item \textit{Cassini/Huygens} sent in 1997 ---terminated in 2017--- by NASA, ESA and the Italian Space Agency (ASI) to study Saturn, its rings and its countless moons, and which stepped on Titan; 
\item \textit{Rosetta/Philae} sent in 2004 ---terminated in 2016--- by ESA which landed on the 67P/Churyumov-Gerasimenko comet; 
\item \textit{Venus Express} sent in 2005 ---terminated in 2015--- by ESA to explore Venus; 
\item \textit{New Horizon} sent in 2006 by NASA to observe the dwarf planet Pluto~; a flyby of one Kuiper belt object is scheduled for 2019;
\item \textit{MAVEN} sent in 2013 by NASA to study the Mars atmosphere; 
\item \textit{OSIRIS-REx} sent in 2016 by NASA to bring samples from an asteroid; 
\item \textit{BepiColombo} scheduled for 2018 by ESA and the Japan Aerospace Exploration Agency (JAXA formely NASDA) to observe Mercury; 
\item \textit{Solar Orbiter} scheduled for 2019 by ESA and NASA to investigate the heliosphere and solar wind. 
\end{itemize}

The \textit{Hubble Space Telescope} is one of the most precious tools for astronomers since 1990, because it is not affected by the atmosphere. Developed by NASA and ESA, it has largely contributed to increase our knowledge on lots of domains in astrophysics and cosmology, like on our solar system, stellar evolutions, interstellar medium, far away galaxies, exoplanets, the supermassive black hole at the center of our galaxy, or the accelerated expansion of the universe.
\textit{Hubble} is orbiting at LEO height (600~km) and uses solid-state amplifiers. But its successor, the \textit{James-Webb Space Telescope} (\textit{JWST} or \textit{Webb}) scheduled for launch in 2018, and developed by NASA, ESA and the Canadian Space Agency (CSA), will use Ka-Band traveling-wave tubes from TED.
\textit{Webb} will be positioned at about 151.1~million kilometres from us, in a Sun--Earth Lagrangian point, a privileged spot for observation. If we compare the contribution of \textit{Hubble}, with its 2.5~meter wide mirror, with promises of \textit{Webb} and its mirror of 6.5~meter, we cannot conceive yet the tremendous scientific contribution which this new satellite will bring, transmitted to Earth thanks to TWTs.

Because of the Earth curvature, keeping a continuous link between objects at LEO and their operation centres is not possible. During the 1980s, NASA started to launch the Tracking and Data Relay Satellite (\textit{TDRS}) system, a constellation of ten GSO satellites ---the last one, \textit{TDRS-M}, was sent on August~18, 2017--- assembled with Ku-Band TWTs, and allowing a permanent coverage for selected missions at LEO. Main representative missions  are the \textit{Hubble Space Telescope} and the \textit{International Space Station} (\textit{ISS}), both equipped with solid-state amplifiers, and before them, the system was operating for the Space Shuttle program. \textit{TDRS} satellites act as relay satellites, always keeping the space-to-ground connection with stations. The rock star of this system was \textit{TDRS-1}, launched on April 1983, which was the first satellite able to see both Poles simultaneously, performing the first Pole-to-Pole call, and was providing most of the \textit{TDRS} coverage\footnote{The second TDRS satellite was destroyed on January~28, 1986 in the Space Shuttle \textit{Challenger} explosion.}. Originally planned for a duration of 7~years, \textit{TDRS-1} was in operation for 27~years, until its last traveling-wave tube failed \cite{zal11}, leaving the craft unable to operate its retransmission activities and expediting its retirement.

\bigskip

Most of recent LEO constellations do not use any TWTs. Solid state power amplifiers (SSPAs) grabbed their share of this market since they are cheaper than vacuum electronics.
\textit{OneWeb} constellation (to be launched in a few years) designed with more than eight hundred satellites to global Internet distribution, will work with Ku-Band SSPAs. The first generation of the \textit{O3b} constellation\footnote{The \textit{O3b} constellation was referring to the ``Other 3 billion'' people without Internet access at that time.}, providing Internet between the two tropics, were equipped with Ka-Band TWTs. However, its second generation used SSPAs. But SSPAs cannot yet reach the high power and/or high frequencies available with TWTs, like for most of GSO program and deep space missions with high data traffic required. 
Besides, there still are recent satellite families at LEO using TWTs like Earth observation programs ---like the \textit{Sentinel} missions sent by ESA or the Canadian \textit{Radarsat}\footnote{In addition to TWTs for communication, \textit{Radarsat-2} is equipped with enhanced interaction klystrons (EIK) for tomography.} made by CSA---, meteorological programs ---like ESA's \textit{MetOp} satellites---, or the \textit{CryoSat} program of ESA to measure the thickness of polar ices.

\bigskip

Besides, almost all ground stations in contact with spacecrafts use TWTs or klystrons. 

\section*{Conception of traveling-wave tubes}

There exist numerous amplifiers from vacuum tubes, like triodes, magnetrons or klystrons, to solid-state power amplifiers (SSPAs), like transistors. A popular misconception claims tubes to be under threat of extinction for the benefit of solid-state electronics. 
While this is true in computers and modern domestic electronics ---except for microwave ovens---, this is plainly false in space applications. 
TWTs provide a crucial service to space telecommunication and have continuously been improved.
There is no perfect amplifier for all needed applications, but there are better ones depending on their performances, their cost and their operating regimes, and traveling-wave-tube amplifiers still dominate satellite communications. 

The number of space TWTs produced each year depends on the number of satellites launched. Since 2010, there were approximatively twenty GSO satellites sent per year, and a communication satellite contains between fifty to sixty amplifiers. Consequently, there are between five hundred to two thousand space TWTs produced per year. The first manufacturer is Thales Electron Devices (TED) with two-thirds of the market share, followed by L3 Electron Devices with the remaining third.

\bigskip

We believe that the traveling-wave tube can be considered as the purest microwave device due to its harmonic way of using the wave-particle interaction (see appendix~A). It is also probably the hardest vacuum tube to build due to the Swiss watch precision required for highly rugged elements.
The choice of an amplifier is determined by four main characteristics (in importance order): reliability, performance, weight, and price.
Current space TWTs prices range from $40\,000$ to $150\,000$ euros depending on the precision of the device ---high frequencies need smaller slow-wave structures.
Production of traveling-wave tubes requires handicraft methods.
Manufacturing is divided in two parts separated by the device pumping. The upstream part is the assembly of the diverse components and takes one to two months. Then, an ultra high vacuum ---beyond $10^{-8}$ Pa--- is obtained with pumps while heating tubes to over 500$^{\circ}$C. Finally, the downstream part consists in adjustment to fix the device ---all TWTs are unique---, measurements to know characteristics of each tube, and finally, a burn-in process where devices are tested under various conditions ---like vibrations, vacuum environment, or long runs of functioning. This part takes five to six months. 

\bigskip

The traveling-wave tube is still the most efficient device when one wants to reach high power and/or high efficiency in space telecom. 
For the moment, it is more expensive to use SSPAs above the Ku-Band instead of TWTs. Of course, in the future, development of better SSPAs\footnote{Space SSPAs have some advantage, like a noise factor around 5~dB, compared to 30~dB for space TWTs.} will border the separation between the two, but vacuum valves, especially TWTs, have not yet reached the limit of their improvements.
Several paths for optimization can be taken considering the three parts of TWTs.

As presented by tables  \ref{tab:2}, \ref{tab:3} and \ref{tab:5}, TWTs have been dynamically improved over the years (see \cite{kor01}). Compared to their historical debut, engineers have increased the number of collectors to three, then four, and now five, with a massive progression of the efficiency. In addition, thanks to the invention of radiation cooling, TWTs are probably the only electronic component operating up to 200$^{\circ}$C. And, the linearized TWTs, which appeared in the 1990s, fully compensate for the tube higher non-linearity.
Also, the power flexibility of the device allows remote control of the output power from the ground.

% Par rapport au début historique on a augmenté le nombre de collecteur à 3, puis 4 et maintenant 5 étages avec une grosse progression sur le rendement. J'ai tenté un graphique sur le rendement à 20 GHz depuis 1995. Nous avons aussi inventé le refroidissement par rayonnement qui fait que le TOP est (probablement) le seul composant électronique qui fonctionne à 200 degC. Un progrès important est aussi venu des linéariseurs dans les années 1990. On pourrait enfin mentionner la flexibilité en puissance qui permet de télécommander la puissance du TOP en orbite.

How the electron gun injects exactly the electron beam is still under investigation.
The number of particles imposes the use of powerful computers.
The type of slow-wave structures can also be improved (including materials used) to optimize the momentum conversion from electrons to the electromagnetic wave. Indeed, only a quarter of the beam power is transferred.
The rest of it needs to be recovered, lest it will become unwanted thermal power ---heating a TWT modifies its performances.
So the last part to be improved is the depressed collector that determines the efficiency of the TWT.
In addition, a better understanding of nonlinear effects occurring with high amplifications is required to maximize the interaction and to grasp the wave output modulation ---the most important factor in current complex signal transmissions.

\section*{Other applications of traveling-wave tubes}

Furthermore, traveling-wave tubes have other applications than space ones. In addition to ground-to-space transmissions, the device still has some use in telecommunication, including transponders for high frequency military bands. Also, TWTs can be found in command-guidance systems for missiles and in electronic countermeasures (ECM). While this article is focused on its amplification role, the traveling-wave tube can also be used as a radar transmitter: the role for which Kompfner worked on valves.
For those applications, the slow-wave structure can be replaced with coupled cavities or folded wave-guides, and the tube can be a few metres long.

Engineers in microwave radio-relays (still used nowadays) replaced TWTs in stations built from the 1950s to the 1990s, by solid-state power amplifiers over time. Simultaneously, those relays faced the competition with satellite communications, improved cables, and recent optical fibres. But tubes are not gone in data transfers since TWTs can reach higher frequencies.
%\footnote{Even if, nowadays, the faster way to sent a huge amount of data is through mail.}.
For the new 5\textsuperscript{th} generation mobile networks, abbreviated 5G, the \textsc{tweether} project is funded by the European Union, aiming to create state-of-the-art W-Band TWTs, based on a folded wave-guide slow-wave structure. In addition, several research groups aim to build terahertz TWTs (the record is probably 1.03~THz so far \cite{tuc16}).

%TWTs sells themselves like hot cakes.
TWTs represent an important part of the vacuum devices market.
In 1995, the total worldwide sales of all microwave tubes %(including radar, electronic warfare, space, communications, industrial heating, medical and scientific, but excluding equipments to the general public like magnetrons for microwave ovens) 
were estimated at 780 million dollars (taking into account inflation), and over half of it were for TWTs (410 million dollars) \cite{dod97}. This was, and still is, due to the price of TWTs, from a few thousand to several hundred thousand dollars.
In comparison, simple low power magnetrons, like those in microwave ovens for the general public, are relatively inexpensive (a few dollars).

\bigskip

Since the beginning of the space race, a large number of objects are orbiting for centuries in our close neighbourhood. Collisions between them have already occurred, creating large fields of debris, which might make it impossible to achieve any new space mission.
For example, on August 25, 2017, the Indonesian satellite \textit{Telkom~1}, orbiting at GSO, was struck by an unknown object generating a large cloud of debris. This incident caused the breakdown of $12\,000$ cash dispensers in the Indonesian archipelago \cite{unk17}.
To track the hundreds of thousand crafts and debris in orbit, space agencies have started numerous programs to catalogue them to keep space clean.
As an example, ESA collaborates with the Research Establishment for Applied Science (FGAN), Wachtberg, Germany, which have the Tracking and Imaging Radar (TIRA) facility. This installation possesses a high-resolution Ku-Band imaging radar, including a traveling-wave tube, to photograph large objects and verify their integrity \cite{meh02}.
%of operational procedures, attitude determination, emergency operations, damage and fragmentation analysis. 

Traveling-wave tubes have also a role in plasma physics. A plasma is a noisy medium and is hard to control and to analyse. Researchers at the University of California, San Diego, noted that a helix slow-wave structure can behave analogously to a plasma. To study self-consistent and nonlinear effects (e.g.\ trapping) of electrons in plasma waves, they built \cite{dim77,tsu87} in 1976 a huge TWT, 2.7 meters long, with 2.1~centimetres diameter and 1~millimetre pitch for frequencies around 50~MHz. 
Later, a team from CNRS \& Aix-Marseille University, brought forward the idea, this time to investigate the chaos transition occurring in plasma physics \cite{guy96,Dov07}. Their tube, still used, is one of the longest TWTs for civil use in the world with its 4~meters ---long enough to need magnets to compensate for the Earth magnetic field.

%\bigskip
%
%Other vacuum electron devices are still used.
%Klystrons are used as wave amplifiers on radars and particle accelerators, like for radiotherapy. Gyrotrons are a major source for fusion plasma heating in magnetic confinement devices.
%Finally, the physics underlying TWTs can be used to understand Linear Accelerators (LINACs) and Free Electron Lasers (FELs).

\section*{Conclusion}

The same year \textit{Telstar~1} reached posterity, Clarke claimed, with his Third Law, that ``any sufficiently advanced technology is indistinguishable from magic'' \cite{Cla73}. If long-distance communication seemed impossible a hundred years ago, the traveling-wave tube made it possible from outer space.

The traveling-wave tube is one of the most appropriate vacuum electron tubes in the field of telecommunication.
Conceiving one requires the precision of a goldsmith and to appreciate the exchange of momentum between electromagnetic waves and electrons (see Appendix~A). Relaying on decades of progress in delay-lines and thermionic valves, Haeff, Lindenblad and Kompfner were successively able to understand the intuitive basis of this process. They imagined a device where the electron beam accompanied the wave by making their velocities almost equal.

Since then, TWTs appeared on the majority of spacecrafts; from LEO research probes, to deep space missions, and GSO communication satellites.
Numerous people and organizations took part in their development and expansion. 
The tube is still at the cutting edge of space power amplifiers and will certainly continue for several decades ; their remote future will depend on further researches to enhance its competitiveness.
As the challenges continue to grow with space exploration reaching always further, or Internet creating always more data, the research and industrial communities are still very active to improve them.

\bigskip

But back in the 1950s, a lot of vacuum tube designs were available. Yet, in less than two decades, the TWT was on every continent and even in space. Its success was due to many factors. The early researches, lead by Kompfner and Pierce, showed encouraging results and were announced at popular conferences and journals among electrical engineers. The devices they built were able to generate waves at high power and high frequencies, but mainly they had large bandwidths for low noise levels, making TWTs a perfect choice for the new requests on communication systems.
Consequently, a lot of competitors appeared, allowing a quick development of the device through all of its aspects. Most important, the device appeared in peace time, at the time when people were interested by the emergence of the television.
Between the 1950s and the 1980s, the appearance of microwave radio-relays, made thanks to TWTs, created one of the first public global networks across continents.
% Indeed, TWTs were highly suitable for telecommunication systems, and still are affordable for satellite contacts and this will certainly continue for several decades.

Extension of this network was possible via space. During the end of the 1950s, and after attempts at using the Moon as a relay, Clarke, Pierce, and Rosen put forward fundamental ideas to develop communication satellites.
The traveling-wave tube was certainly a part of the success, since it allowed satellites to be cost-effective. Successive accomplishments of \textit{Telstar~1}, \textit{Relay~1}, and \textit{Syncom~2} and \textit{3}, proved the considerable advantage of space communications and drew mandkind into a new era.

\appendix

\section*{Appendix A: The wave-particle interaction} 
\label{AppendixA}

Predictions about wave-matter interactions occurring in traveling-wave tubes are made with the same tools as to study plasma evolutions, because the electron beam is a plasma itself with a single species. Basically, there are three model classes\footnote{Quantum effects are not significant in space TWTs since electron energies are not high enough.}: fluid, kinetic (like Vlasovian description) and $N$-body descriptions, each with a wide range of different models and variations.
We choose to focus on the latter like \cite{els03} from a hamiltonian approach.
In the particle description (a.k.a.\ discrete particle or finite $N$ description), each particle of the system is governed by elementary mechanics (like the Lorentz force). Discrete $N$-body description for $N \rightarrow \infty$ become continuous distribution functions within the kinetic picture \cite{els14}.
%Continuous distribution functions from the kinetic description become discrete within the $N$-body picture.
The $N$-body description is hardly ever used to describe the beam in TWTs for it involves an enormous number of degrees of freedom, which increases its computing costs and running time excessively.
However, this description\footnote{The particle description is common in Free Electron Laser (FEL) characterization (see \cite{bon90} for such description and \cite{pel12} for a historical review on FELs). It is also used to study Hamiltonian chaos (see \cite{esc10} and the historical review \cite{esc17}) and the laser-plasma interaction (see \cite{ben16}).} is a remarkably intuitive way to represent wave-particle dynamics.

\bigskip

\begin{figure}[!t]
\centering
\includegraphics[width=0.7\textwidth]{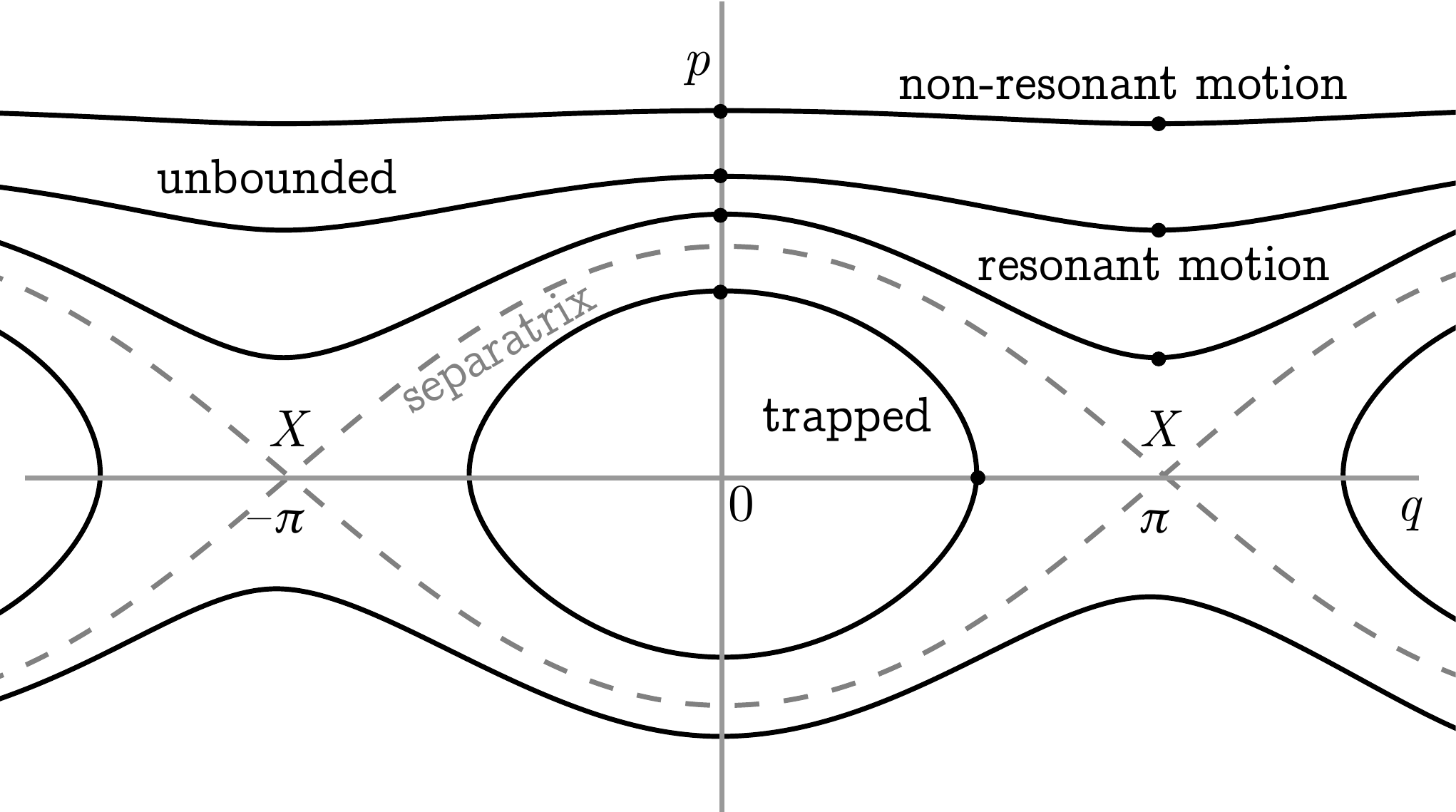}
\caption{\label{f.PhaseSpaceWaticule} Pendulum phase portrait $(p,q)$ of the wave-electron interaction. The width of the cat's eye enclosed by the separatrix (dashed curve) is proportional to $\sqrt{U}$. This portrait is in the reference frame of the wave, so $p=0$ correspond to synchronism with the wave. The further a particle momentum is from $p=0$, the less its dynamics is affected. A resonant motion occurs when the particle dynamics is significantly affected by the wave (when both wave and particles velocities match), viz.\ in the range $|p| \lesssim 3 \sqrt{U}$.}
\end{figure}

We start from a system composed of a single charged particle with coordinate $z=q/k$ and momentum $p$, subject to a sinusoidal electrostatic potential\footnote{
Indeed, we consider an electron beam with initial speed $v_{\mathrm{el},0}$ near resonance with a sinusoidal electric field $E_z (z,t) = E_{z,0} \sin(k z - \omega t)$ projected on the $z$-axis of the beam, with a phase velocity $v_{\mathrm{ph}} = \omega / k$, such that $v_{\mathrm{el},0} \simeq v_{\mathrm{ph}}$.
From Newton's second law, we write the equation of motion of each electron ${\melec} \ddot{z} = - |e| E_z$, or
\begin{equation}
\ddot{z}' = \frac{- |e| E_{z,0}}{\melec} \sin(k z') \,  ,\label{e:NewtonElec}
\end{equation}
after the substitution $z' = z - v_{\mathrm{ph}} t$ to the reference frame where the wave is immobile, with $|e|$ the elementary electric charge and $\melec$ the electron mass.
Equation \eqref{e:NewtonElec} is the same as eq.\ \eqref{e:MvtResonance} with $U = |e| E_{z,0} / \melec$, leading to the same phase portrait as figure \ref{f.PhaseSpaceWaticule}.
Acceleration areas are where $kz' \in [-\pi  + 2 n \pi , 0+ 2 n \pi ]$, with $n\in \ZZ$, and  deceleration areas range where $kz' \in [ 0+ 2 n \pi , \pi  + 2 n \pi ]$. 
This effect causes a particle grouping, called electron bunching.}
with wavenumber $k$. The equation of motion of this particle reads
\begin{equation} \label{e:MvtResonance}
\ddot{q} = - k U \sin q \, ,
\end{equation}
with $U>0$. This system obeys the same mathematical relation as the nonlinear pendulum if we let $k U=g/l$, with $g$ the gravity acceleration, $l$ the length of the pendulum and $q$ the angle with the vertical axis. 

\begin{figure}[!t]
\centering
\includegraphics[width=\textwidth]{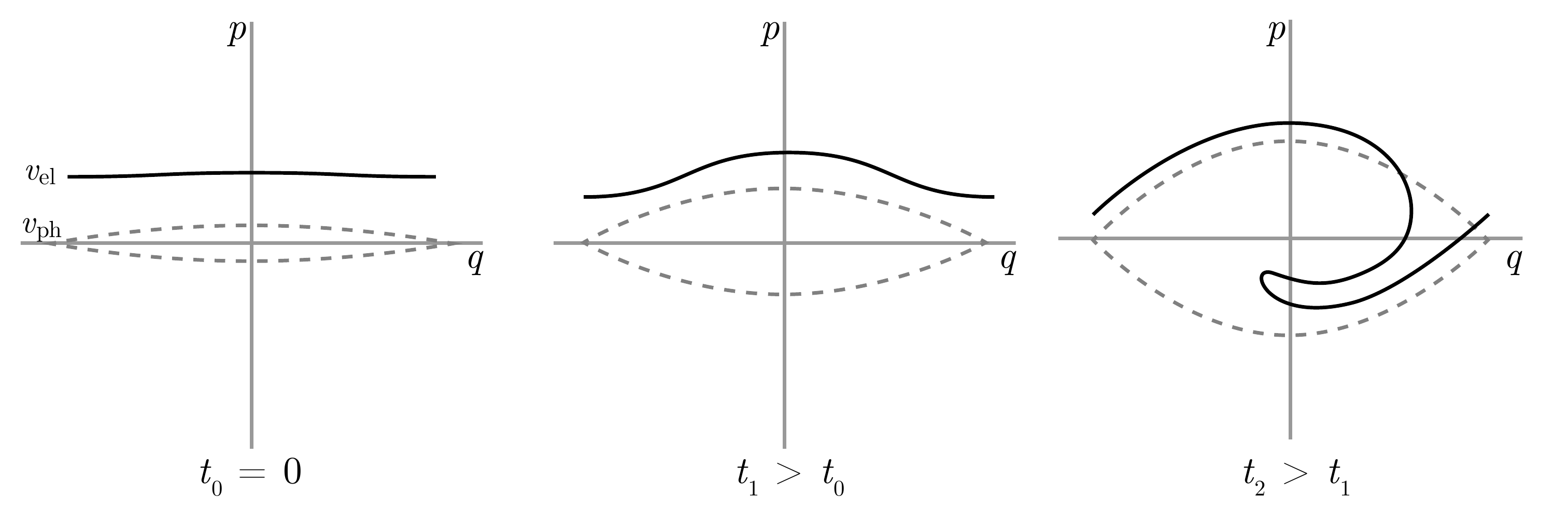}
\caption{\label{f.PhaseSpaceAmpli} Wave-electron momentum exchange for a self-consistent system, inspired from \cite{fir01}.
The width of the cat's eye (dashed lines) is proportional to $\sqrt{U} \sim I^{1/4}$. 
At time $t_0$ (left panel), we consider a monokinetic beam of speed $v_{\mathrm{el}}$ where particles are represented by the continuous black curve. 
We add a wave of phase velocity $v_{\mathrm{ph}} \lesssim v_{\mathrm{el}}$, and trajectories get modulated. 
Some particles get accelerated by the wave, and others get decelerated ; 
it turns out that the deceleration effect dominates the acceleration, so that the total momentum of particles decreases. 
This lost momentum is transferred to the wave, whose 
%we consider all particles already trapped inside the separatrix (nonlinear regime).  At time $t_1 \geq t_0$, each particle has followed its orbits in its $(p,q)$ space. Since they have lost their momenta (except two), relation \eqref{e:momenta} implies that the missing momentum is transferred to the wave, so the wave 
amplitude (hence the cat's eye size) increases. At later times, resp.\ $t_1$ and $t_2$, the wave is amplified (central panel), 
and this process goes on until particles get trapped, which initiates the nonlinear regime due to bunching (right panel). 
The reverse effect occurs if we take $v_{\mathrm{ph}} \gtrsim v_{\mathrm{el}}$: particles will gain momentum and the wave amplitude will decrease \cite{dov05,esc10}. Remark that this example expressed the amplification in time whereas the amplification of a TWT is in space (see figure \ref{f.vElecTWTDIMOHA}). For the traveling-wave tube, nonlinearity effects (as at $t_2$) have been studied since the early years \cite{Cut56}.}
\end{figure}
 
The kinetic energy of this particle, with its mass set to 1, is $p^2/2$~; its potential energy is $- U \cos q$, 
and the sum of both is the total energy (namely, the hamiltonian)
\begin{equation}\label{e:HamilPendul}
H(p,q) = \frac{p^2}{2} - U \cos q \, .
\end{equation}
Since this hamiltonian is time-independent, it is also a constant of motion in phase space $(p,q)$.
Figure \ref{f.PhaseSpaceWaticule} shows typical orbits for the particle trajectory depending on the value of $H$. If the particle is located between $q=0$ and $q=\pi$, it faces a potential barrier at $\pi$: it is slowed-down because of the effect of the field. If the particle is located between $q=-\pi$ and $q=0$, it faces a potential well at $0$ and is accelerated. Depending on the value of $H$, we have
\begin{itemize}
\item[---] For $H > U$, the particle is not confined spatially and reaches all possible positions $q$. It is just like the pendulum in perpetual rotation in one direction (forward for $p > 0$, backward for $p < 0$). Then there are two possible cases. If $H\gg U$, the particle does not ``see'' the field, and its momentum is barely modified. If $H \gtrsim U$, the particle is near resonance with the field: its momentum is modulated depending on the wave.
\item[---] For $-U < H < U$, the particle is trapped in the potential well and traces a closed curve in phase space which never reaches $-\pi$ and $\pi$ positions. This corresponds to the pendulum librating between two points, without having enough energy to complete a full turn. Again, one may distinguish between $1 - H/U \ll 1$, in which case the particle is close to the separatrix, undergoes large momentum changes, and spends much time close to the $X$ point, and the case $1 - H/U \sim O(1)$ in which the particle is ``deeply trapped'' and oscillates mildly around the $O$ point.
\item[---] For $H = U$, the particle is located on the separatrix which is made of two unstable orbits forming a ``cat's eye'' in phase space between libration and rotation motions.
\end{itemize}
However, this model is incomplete. Indeed, its neglects the wave alteration by the presence of charged particles.

\begin{figure}[!t]
\centering
\includegraphics[width=\textwidth]{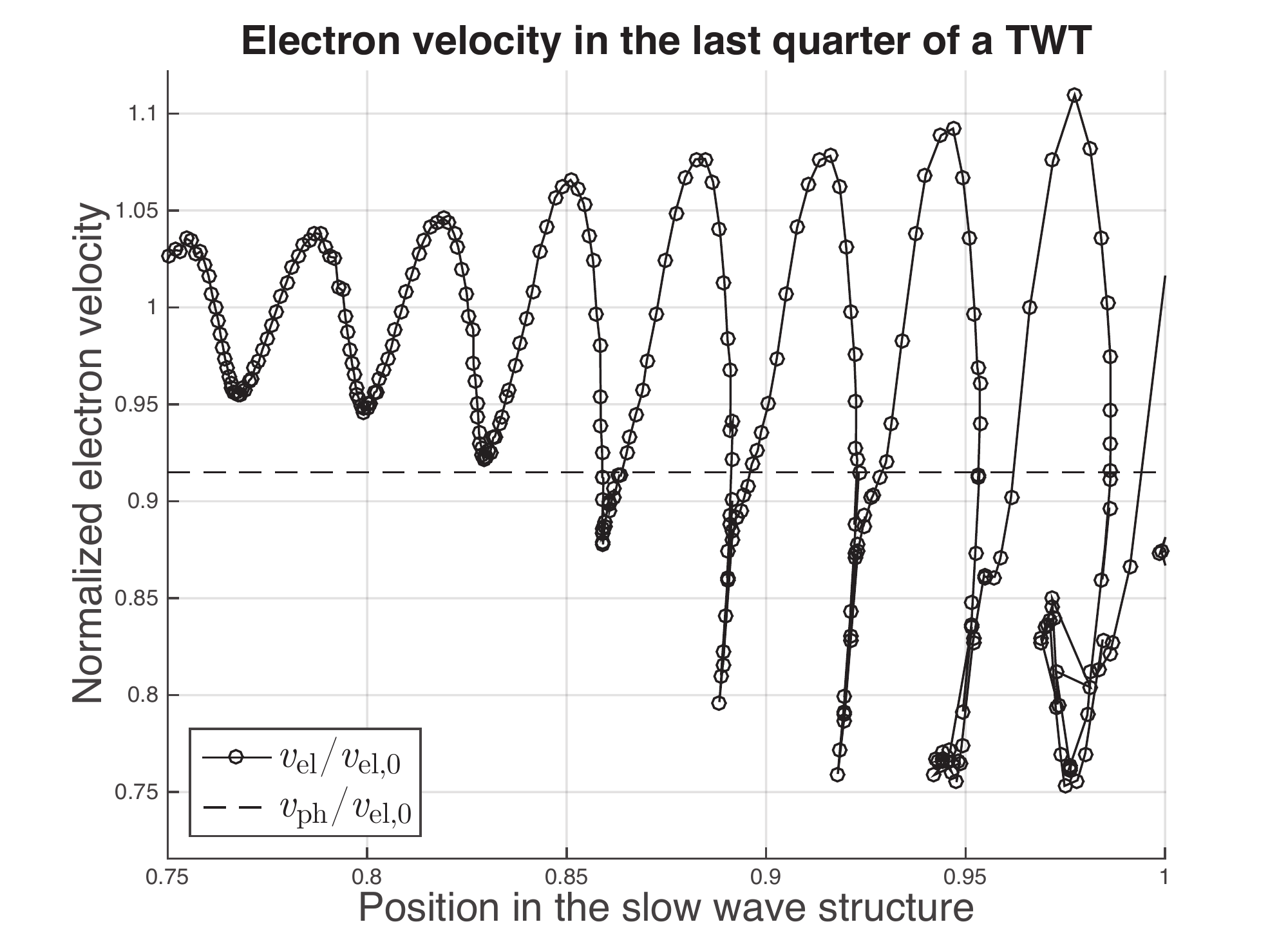}
\caption{\label{f.vElecTWTDIMOHA} One-dimensional time simulation of a traveling wave tube (excluding tapers or attenuations) from a self-consistent hamiltonian model \cite{min17a}. Initial parameters (like cathode current and potential, coupling impedances, and phase velocity) are set to ensure the amplification to reach the power saturation at the end of the tube ($z=1$), and space charge effects are taken into account. Each dot correspond to a macro-electron (with charge $|Q| \sim 200 \, 000 \, |e|$). At the beginning of the tube ($z=0$), all particles are emitted with the same velocity $v_{\mathrm{el},0}$. Particles with $v_{\mathrm{el}}< v_{\mathrm{el},0}$ have given their momenta to the wave, and since there are more particles below $v_{\mathrm{el}} = v_{\mathrm{el},0}$, then the wave is amplified. The dashed line represent the phase velocity $v_{\mathrm{ph}}$ of the wave. Nonlinear effects (trapping) start to occur approximatively when $z=0.8$ (some particles crossed the separatrix centred on $v_{\mathrm{ph}}$).}
\end{figure}

A model taking into account the action of the wave on electrons and the feedback from electrons to the wave, is called self-consistent. 
When adding the energy of harmonic oscillators corresponding to the free oscillation of the wave, one finds a self-consistent hamiltonian like
\begin{equation}
  \label{e:HamilSC}
H(p,z,I,\varphi) = \sum_r \frac{p^2_r}{2} + \sum_j \omega_j I_j - \sum_j \sum_r U_j \cos (k_j z_r - \varphi_j) \, ,
\end{equation}
for waves ($j$) and particles ($r$), where the electrostatic energy of a wave is
proportional to the square of its amplitude $U_j \propto \sqrt{I_j}$, and waves have phases $\varphi_j$, nominal angular frequencies $\omega_j = v_{\mathrm{ph},j}  k_j$, phase velocities $v_{\mathrm{ph},j}$ and wavenumbers $k_j$. 
The abrupt transition between relations \eqref{e:HamilPendul} and \eqref{e:HamilSC} is further detailed in \cite{ant98,els03}.
As the hamiltonian \eqref{e:HamilSC} is invariant under time translations ($t' := t + b$), the dynamics conserves total energy $H$. 
Since \eqref{e:HamilSC} is also invariant under space translations ($z'_r := z_r + c$, $\varphi'_j := \varphi_j + k_j c$), 
the dynamics also conserves total momentum 
\begin{equation} 
  \label{e:momenta}
P = \sum_r p_r + \sum_j k_j I_j \, ,
\end{equation}
which is a simple sum of particle and wave momenta. The wave-electron interaction is an effect based on momentum exchange. 
This effect is sketched on figure \ref{f.PhaseSpaceAmpli}.

\bigskip

The physical process in traveling-wave tubes is easy to appreciate (see figure~\ref{f.vElecTWTDIMOHA}). 
The slow-wave structure of the tube is designed to impose quasi-resonance 
between the phase velocity of the wave $v_{\mathrm{ph}}$ along the propagation axis, and the electron speed $v_{\mathrm{el}}$ from the electron gun. 
As shown on figure \ref{f:helix}, the phase velocity in helix tubes is $v_{\mathrm{ph}} \simeq  2 \pi \, c \,  a_{\varnothing} / d$, with the light celerity $c$.
To obtain amplification of the wave, one just needs to ensure $v_{\mathrm{el}} \gtrsim v_{\mathrm{ph}}$. 
During the interaction in the tube, the total momentum \eqref{e:momenta} is conserved, so electrons on average will lose their momentum for the benefit of the wave. This effect is linear (because the amplification gain is linear) until particles have lost enough momentum 
for crossing the separatrix in their individual $(z, p)$ space and become trapped.
More realistic models can contain in particular the space charge effect (viz.\ the Coulombian effect): 
electrons repel each other since they have the same charge sign. 
This effect is noticeable in TWTs, since it reduces the electron bunching and requires slightly longer slow-wave structures. 
We also remark that in real TWTs, emitted electrons are not forming a perfect thin and monokinetic beam, as well as the phase velocity is not constant and losses occur.

%\begin{table}[h]
%\centering
%\caption{ \red{finir}}
%\label{tab:6}       % Give a unique label
%\begin{tabular}{ll}
%\hline\noalign{\smallskip}
%1888 & Discovery of electromagnetic waves in the air by Hertz \\
%1889 & First slow-wave structure by Hertz \\ 
%1890s & First telegraphic and voice radios \\
%1904 & First vacuum electron device by Fleming \\
%1933 & First traveling-wave tube by Haeff \\
%1940 & Lindenblad's TWT \\
%1942 & Kompfner's TWT \\
%1946 & General announcement of TWT first capacities \\
%1952 & Manchester--Edinburgh: first microwave radio-relay equipped with TWTs \\
%1950s & Early worldwide microwave radio-relays \\
%1962 & \textit{Telstar 1}: first TWT in space \\
%1963 & \textit{Syncom 2}: first TWT in GSO \\
%\end{tabular}
%\end{table}

%\goodbreak
\section*{Further reading}
\begin{itemize}
\item Invention of TWTs \cite{wat54,kom64,kom76,Cop15a}.
\item Technical basis about TWTs \cite{pie50,gil94,fai08}.
\item About space TWTs \cite{kor01}.
%\item Review of space TWTs before 1980 \cite{cuc81,kos83}.
\item History of the space conquest \cite{log95,But97}.
\item History of telecommunication \cite{Bra95,gav98}.
%\item History of Bell Labs \cite{ger12}.
%\item Further details about space missions \cite{ESA,NASA}.
\end{itemize}

\section*{Acknowledgements}

It is a pleasure for the authors to gratefully acknowledge Olivier Agullo, Dominique Escande and Patricia Radelet-de Grave for their critical reading of the manuscript, and Guillaume Fuhr and Magali Muraglia for their stimulating comments.

\end{document}